\documentclass[journal]{IEEEtran}
\IEEEoverridecommandlockouts

\usepackage{amsthm}

\newtheorem{thm}{Theorem}[section]
\newtheorem{lem}{Lemma}[section]

\newtheorem{cor}{Corollary}[section]

\theoremstyle{definition}
\newtheorem{defn}{Definition}[section]
\newtheorem*{defn*}{Definition}
\newtheorem{scheme}{Scheme}[section]
\newtheorem*{scheme*}{Scheme}

\theoremstyle{remark}
\newtheorem{remark}{Remark}[section]

\usepackage[tbtags]{amsmath} 
\usepackage{amssymb}  
\usepackage{verbatim} 
\usepackage{amsxtra}  
\usepackage{comment}

\usepackage{mathabx}

\usepackage{enumerate}

\usepackage{amsfonts}
\usepackage{color}

\usepackage{subcaption}

\usepackage{wasysym}

\usepackage{cite}

\usepackage{tikz}
\usetikzlibrary{shapes,arrows,decorations.markings,positioning,calc}

\tikzstyle{plant} = [draw, fill=red!5, rectangle, 
    minimum height=3em, minimum width=6em]
\tikzstyle{block} = [draw, fill=blue!5, rectangle, 
    minimum height=3em, minimum width=18mm]
\tikzstyle{sum} = [draw, fill=yellow!10, circle, node distance=1cm]
\tikzstyle{coord} = [coordinate]
\tikzstyle{gain} = [draw, fill=red!5, regular polygon, regular polygon sides=3, shape border rotate=-90]
\tikzstyle{pinstyle} = [pin edge={to-,thick,black}]

\tikzstyle{BitPipe} = [thick, decoration={markings,mark=at position
   1 with {\arrow[semithick]{open triangle 60}}},
   double distance=1.4pt, shorten >= 5.5pt,
   preaction = {decorate},
   postaction = {draw,line width=1.4pt, white,shorten >= 4.5pt}]

\usetikzlibrary{fit,backgrounds}

\usepackage{epsfig, graphicx, psfrag}

\usepackage[mathcal]{euscript}
\usepackage[cal=boondox]{mathalfa}

\providecommand{\thmref}[1]{Th.~\ref{#1}}

\providecommand{\defnref}[1]{Def.~\ref{#1}}
\providecommand{\secref}[1]{Sec.~\ref{#1}}
\providecommand{\lemref}[1]{Lem.~\ref{#1}}

\providecommand{\remref}[1]{Rem.~\ref{#1}}
\providecommand{\figref}[1]{Fig.~\ref{#1}}
\providecommand{\corref}[1]{Cor.~\ref{#1}}
\providecommand{\appref}[1]{App.~\ref{#1}}

\providecommand{\schemeref}[1]{Sch.~\ref{#1}}

\usepackage{refcount}

\newcommand{\iid}{i.i.d.}

\newcommand{\viz}{viz.}
\newcommand{\etal}{\emph{et al.}}

\newcommand{\reals}{\mathbb{R}}
\newcommand{\ints}{\mathbb{Z}}

\newcommand{\nats}{\mathbb{N}}

\DeclareMathOperator*{\argmax}{argmax}

\newcommand{\SNR}{\mathrm{SNR}}
\newcommand{\ENR}{\mathrm{ENR}}

\newcommand{\SDR}{\mathrm{SDR}}

\newcommand{\eff}{{\mathrm{eff}}}

\newcommand{\e}{\mathrm{e}}

\newcommand{\mX}{\mathcal{X}}

\newcommand{\mF}{\mathcal{F}}
\newcommand{\mH}{H}

\newcommand{\mV}{\mathcal{V}}

\providecommand{\hmX}{\hat{\mX}}

\newcommand{\Comment}[1]{}
\newcommand{\old}[1]{}
\newcommand{\rem}[1]{}

\newcommand{\eps}{{\epsilon}}
\newcommand{\teps}{{\tilde{\eps}}}
\newcommand{\veps}{\varepsilon}

\newcommand{\err}{\mathrm{err}}
\newcommand{\MMSE}{{\mathrm{MMSE}}}

\newcommand{\hx}{{\hat{x}}}
\newcommand{\hm}{\hat{m}}

\newcommand{\htm}{\hat{\tilde{m}}}

\newcommand{\tm}{{\tilde{m}}}
\newcommand{\tL}{{\tilde L}}

\newcommand{\tC}{\tilde C}

\newcommand{\tE}{\tilde{E}}

\newcommand{\tn}{\tilde{n}}

\newcommand{\tD}{\tilde D}

\newcommand{\tH}{\tilde H}

\newcommand{\ty}{\tilde y}

\newcommand{\talpha}{{\tilde \alpha}}

\newcommand{\abs}[1]{\left| #1 \right|}
\newcommand{\Norm}[1]{\left\| #1 \right\|}

\providecommand{\e}{{\rm e}}
\providecommand{\comment}[1]{}

\providecommand{\norm}[1]{\Norm{#1}}

\newcommand{\beqn}[1]{\begin{eqnarray}\label{#1}}
\newcommand{\eeqn}{\end{eqnarray}}
\newcommand{\beq}[1]{\begin{equation}\label{#1}}
\newcommand{\eeq}{\end{equation}}

\newcommand{\MAP}{\mathrm{MAP}}

\providecommand{\half}{\frac{1}{2}}

\providecommand{\half}{\frac{1}{2}}

\makeatletter
\newcommand{\vast}{\bBigg@{4}}
\newcommand{\Vast}{\bBigg@{5}}
\makeatother

\providecommand{\bbE}{\mathbb{E}}

\providecommand{\E}[1]{\bbE \left[ #1 \right]}
\providecommand{\CE}[2]{\bbE \left[ #1 \middle| #2 \right]}

\renewcommand{\gamma}{\ENR}

\providecommand{\Edsgn}{\tE}
\newcommand{\PR}[1]{\Pr\left( #1 \right)}

\providecommand{\tsigma}{\tilde{\sigma}}
\providecommand{\Nmin}{N_{\min}}

\providecommand{\bullets}{\bullet \bullet \bullet}

\providecommand{\cov}[2]{\mathrm{cov} \left(#1, #2  \right)}

\usepackage{ifthen}
\usetikzlibrary{circuits.ee.IEC}
\usepackage{circuitikz}

\usepackage{mathtools}

\mathtoolsset{showonlyrefs=true}

\usepackage{MnSymbol}

\usepackage{comment}
\newcommand{\ignore}[1]{}

\newcommand{\VersionLength}{short}

\providecommand{\ver}{\ifthenelse{\equal{\VersionLength}{long}}}
\providecommand{\nver}{\ifthenelse{\equal{\VersionLength}{short}}}

\providecommand{\figref}[1]{Fig.~\ref{#1}}
\providecommand{\secref}[1]{Sec.~\ref{#1}}

\providecommand{\ColumnNum}{2}
\newcommand{\col}{\ifthenelse{\equal{\ColumnNum}{1}}}

\begin{document}
\title{Universal Joint Source--Channel Coding Under an Input Energy Constraint}

\author{Omri Lev\textit{, Graduate Student Member, IEEE}, and Anatoly Khina\textit{, Member, IEEE}
    \thanks{This work was supported by 
    the \textsc{Israel Science Foundation} (grant No.\ 2077/20).
    The work of O.~Lev was further supported by the Yitzhak and Chaya Weinstein Research Institute  for Signal Processing. 
    The work of A.~Khina was further supported by the WIN Consortium through the Israel Ministry of Economy and Industry.
    This work was presented in part at the
   2022 IEEE International Symposium on Information Theory (ISIT), Espoo, Finland.}
    \thanks{O.~Lev was with the School of Electrical Engineering, Tel Aviv University, Tel Aviv~6997801, Israel. He is now with the Signals, Information and Algorithms Laboratory, Massachusetts Institute of Technology (MIT), Cambridge, MA~02139, USA
    (e-mail: \texttt{omrilev@mit.edu}).}
    \thanks{A.~Khina is with the School of Electrical Engineering, Tel Aviv University, Tel Aviv~6997801, Israel (e-mail: \texttt{anatolyk@eng.tau.ac.il}).}
}

\maketitle


\begin{abstract}
    We consider the problem of transmitting a source over an infinite-bandwidth additive white Gaussian noise channel with unknown noise level under an input energy constraint.
    We construct a universal scheme that uses modulo-lattice modulation with multiple layers;
    for each layer, we employ either analog linear modulation or analog pulse position modulation (PPM). 
    We show that the designed scheme with linear layers requires less energy compared to existing solutions to achieve the same quadratically increasing distortion profile with the noise level; replacing the linear layers with PPM layers offers an additional improvement. 
\end{abstract}


\allowdisplaybreaks

\section{Introduction}
\label{s:intro}

Due to the recent technological advancements in sensing technology and the internet of things, 
there is a growing demand for low-energy communications solutions. Indeed, since many of the sensors have only limited battery due to environmental (in case of energy harvesting) or replenishing limitations, these solutions need to be economical in terms of the utilized energy. 
Moreover, since each sensor may serve several parties, with each experiencing different conditions, these solutions need to be robust with respect to the noise level.

This problem may be conveniently modeled as the classical setup of conveying $k$ independent and identically distributed (\iid) source samples over a continuous-time additive white Gaussian noise (AWGN) channel under an energy constraint per source sample. 

In the limit of a large source blocklength, $k \to \infty$, 
and when the noise level is known at both the transmitter and the receiver, 
the optimal performance is known and is dictated by the celebrated source--channel separation principle \cite[Th.~10.4.1]{CoverBook2Edition}, \cite[Ch.~3.9]{ElGamalKimBook}. 
For a memoryless Gaussian source and a quadratic distortion measure, the minimal (optimal) achievable distortion $D$ is given by 
\begin{align}
    D = \sigma_x^2 \cdot \e^{-2\ENR},
\label{eq:SeparationBound}
\end{align}
where $\ENR$ denotes the energy-to-noise ratio (ENR) over the channel, and $\sigma_x^2$ is the source variance.
For other continuous memoryless sources, the optimal distortion is bounded as \cite[Prob.~10.8, Th.~10.4.1]{CoverBook2Edition}, \cite[Prob.~3.18, Ch.~3.9]{ElGamalKimBook}
\begin{align}
    \frac{\e^{2h(x)}}{2\pi\e} \cdot \e^{-2\ENR} \leq D \leq \sigma_x^2 \cdot \e^{-2\ENR} ,
\end{align}
where the lower bound stems from Shannon's lower bound \cite{Shannon59:RDF}, the upper bound holds since a Gaussian source is the ``least compressable'' source with a given variance under a quadratic distortion measure, and $h(x)$ denotes the differential entropy of the source $x$ \cite[Ch.~8]{CoverBook2Edition}, \cite[Ch.~2.2]{ElGamalKimBook}.

While the optimal performance is known when the transmitter and the receiver is cognizant of the noise level and $k \to \infty$, determining it becomes much more challenging when the noise level is unknown at the transmitter.
Indeed, when the transmitter is oblivious of the true noise level achieving \eqref{eq:SeparationBound} for all noise levels simultaneously is not possible \cite{KokenTuncel}. Instead, one wishes to achieve graceful degradation of the distortion with the noise level.\footnote{\label{foot:unlimited-BW}Since the available bandwidth is unlimited, the receiver can learn the white noise level within any accuracy. Moreover, for unlimited bandwidth, the same performance can be attained for any (even infinite) transmission duration.}

For the case of finite bandwidth-expansion/compression $B$ (and finite power),
by superimposing digital successive refinements \cite{EquitzCover91} with a geometric power allocation, 
Santhi and Vardy \cite{Santhi-Vardy:JSCC:optimal-slope:ISIT,Santhi-Vardy:JSCC:optimal-slope:Arxiv}, and Bhattad and Narayanan \cite{Bhattad-Narayanan:JSCC:optimal-slope} showed that the distortion improves $\SNR^{-(B-\eps)}$ for an arbitrarily small $\eps>0$, 
for \textit{large SNR values}. We note that this suggests that, by taking the bandwidth to be large enough,
a polynomial decay with the $\SNR$ of any finite degree, however large, is achievable, starting from a large enough $\SNR$. 
In our setting of interest, this means, in turn, that there exists a finite
energy E for which a polynomial profile 
\begin{subequations} 
\label{eq:profile:tot} 
\begin{align}
\label{eq:profile}
    D &\leq \sigma_x^2 \mF(N) & \forall N > 0
\end{align}
with
\begin{align}
\label{eq:Profile_Polynom} 
    \mF(N) \triangleq \frac{1}{1 + \left(\frac{\Edsgn}{N}\right)^L}    
\end{align}
\end{subequations} 
is attainable for any $1 \leq L < \infty$, however large, 
where $\Edsgn > 0$ is a predesigned normalization constant of our choice.

Mittal and Phamdo \cite{MittalPhamdo} constructed a different scheme that works above a certain minimum (not necessarily large) design signal-to-noise ratio (SNR) by sending the digital successive refinements incrementally over non-overlapping frequency bands, and sending the quantization error of the last digital refinement over the last frequency band. 

The scheme of Mittal and Phamdo was subsequently improved by Reznic \etal~\cite{Reznic} (see also \cite{JointWZ-WDP,AnalogMatching}, \cite[Ch.~11.1]{ZamirBook}), by replacing the successive refinement layers with lattice-based Wyner--Ziv coding \cite{WynerZiv76,Wyner78}, \cite[Ch.~11.3]{ElGamalKimBook} which, in contrast to the digital layers of the scheme of Mittal and Phamdo, enjoys an improvement of each of the layers with the SNR.

Kok\"en and Tuncel \cite{MinimumEnergyBound_Tuncel} adopted the scheme of Mittal and Phamdo to the infinite-bandwidth (and infinite-blocklength) setting. Baniasadi and Tuncel \cite{baniasadi2020minimum} (see also \cite{Baniasadi-Koken-Tuncel:energy-limited-JSCC:universal:IT2022}) further improved this scheme by allowing sending the resulting analog errors of all the digital successive refinements. 
For the case of a distortion profile that improves quadratically with the ENR [$L=2$ in \eqref{eq:profile:tot}] 
upper and lower bounds were established by K\"oken and Tuncel \cite{MinimumEnergyBound_Tuncel} and Baniasadi and Tuncel \cite{baniasadi2020minimum} (see also \cite{Baniasadi-Koken-Tuncel:energy-limited-JSCC:universal:IT2022}) for the minimum required energy to attain such a profile for all ENR values: 
For $\Edsgn > 0$ and a Gaussian source, a quadratic distortion profile \eqref{eq:profile:tot} with $\Edsgn$ (and $L=2$) is achievable with a minimal transmit energy that is bounded as\footnote{More precisely, the achievability results of \cite{MinimumEnergyBound_Tuncel,baniasadi2020minimum} state that for $\Nmin > 0$, however small, the profile \eqref{eq:profile:tot} with $L=2$ and a predefined $\Edsgn$ is achievable for all $N > \Nmin$ for $E = 2.32 \Edsgn$.}
\begin{align}
\label{eq:minimum-energy-bounds}
    0.906 \Edsgn \leq E \leq 2.32 \Edsgn.
\end{align}

Furthermore, K\"oken and Tuncel \cite{MinimumEnergyBound_Tuncel} proved that an exponential profile---\eqref{eq:profile} with $\mF(N) = a \e^{b N}$ for all $N > 0$ for some $a,b > 0$---cannot be attained with finite transmit energy. A staircase profile was treated by Baniasadi \cite{baniasadi202staircase} (see also \cite{Baniasadi-Koken-Tuncel:energy-limited-JSCC:universal:IT2022}).

In this work, we adapt the modulo-lattice modulation (MLM) scheme of Reznic \etal~\cite{Reznic} 
with multiple layers to the infinite-bandwidth setting. 
By utilizing linear modulation for all the layers, we show that this scheme improves the upper (achievability) bound in \eqref{eq:minimum-energy-bounds}. 
Following \cite{EnergyLimitedJSCC:Lev_Khina:Full}, we then replace the analog modulation in (some of) the layers with analog pulse position modulation (PPM). We show that this scheme requires less energy to attain the same quadratic distortion profile compared to the linear layer-only MLM scheme.

The rest of the paper is organized as follows. 
We introduce the notation that is used in this work in \secref{ss:notation}, 
and formulate the problem setup in \secref{s:Problem Statement}. 
We provide the necessary background of MLM and analog PPM in \secref{s: MLM} and \secref{s: AnalogPPM}, respectively.
We then construct universal schemes in \secref{s: UnKnownENR_SchemeDesign}; simulation results are provided in \secref{ss:numeric}. Finally, we conclude the paper with \secref{s:Summary} and \secref{s:FutureResearch} by discussing future research directions and possible improvements.


\subsection{Notation}
\label{ss:notation}

$\nats$, $\reals$, $\reals_+$ denote the sets of the natural, real and the non-negative real numbers, respectively.
With some abuse of notation, 
we denote tuples (column vectors) by $a^{k} \triangleq \left(a_1, \ldots, a_{k} \right)^\dagger$ for $k \in \nats$, and their Euclidean norms---by $\norm{a^k} \triangleq \sqrt{\sum_{i=1}^{k} a_i^2}$, where $(\cdot)^\dagger$ denotes the transpose operation; distinguishing the former notation from the power operation applied to a scalar value will be clear from the context. The i'th element of the vector $a^k$ denoted by $a_i$ or by $a\left[i\right]$, where we will use both notations throughout the paper.
All logarithms are to the natural base and all rates are measured in nats.
The differential entropy of a continuous random with probability density function $f$ is defined by $h\left(x\right) \triangleq -\int_{-\infty}^{\infty}f(x)\log f(x) dx$ and is measured in nats. The expectation of a random variable (RV) $x$ is denoted by $\E{x}$.
We denote by $[a]_L$ the modulo-$L$ operation for $a,L \in \nats$, 
and by $[\cdot]_\Lambda$---the modulo-$\Lambda$ operation \cite[Ch.~2.3]{ZamirBook} for a lattice $\Lambda$ \cite[Ch.~2]{ZamirBook}. $\lfloor \cdot \rfloor$ denotes the floor operation. 
We denote by $I_k$ the $k$-dimensional identity matrix. 
We denote sets of vectors by capital italic letters, where $\mathcal{A}_{b,c}$ stands for a set of $c$ vectors, each of length $b$. 

	
\section{Problem Statement}
\label{s:Problem Statement}

In this section, we formalize the JSCC setting that will be treated in this work.

\textit{Source.}
The source sequence to be conveyed, $x^k \in \reals^k$, comprises $k$ \iid\  samples of a standard Gaussian source.

\textit{Transmitter.}
Maps the source sequence $x^k \triangleq (x_1, x_1, \ldots, x_{k})$ to a continuous input waveform $\left\{ s_{x^k}(t) \bigg{|}  |t| \leq kT/2 \right\}$ that is subject to an energy constraint:\footnote{The introduction of negative time instants yields a non-causal scheme. This scheme can be made causal by introducing a delay of size $kT/2$. We use a symmetric transmission time around zero for convenience.}
\vspace{-.45\baselineskip}
\begin{align}
\label{eq:InputEnergyConstraint}
    \int_{-\frac{kT}{2}}^{\frac{kT}{2}}\abs{s(t)}^2dt &\leq kE &\forall x^k \in \reals^k,
\end{align}
where $E$ denotes the per-symbol transmit-energy.\footnote{$E = PT$ where $P$ is the transmit-power and $T$ is the transmission duration.}

\textit{Channel.}
$s_{x^k}$ is transmitted over a continuous-time additive white Gaussian noise (AWGN) channel:
\begin{align}
 \label{eq:ChannelEq}
     r(t) &= s(t) + n(t), & t \in \left[-\frac{kT}{2},\frac{kT}{2}\right],
\end{align}
where $n$ is a continuous-time AWGN with two-sided spectral density $N/2$,  
and $r$ is the channel output signal; $N$ is referred to as the noise level.

\textit{Receiver.}
Receives the channel output signal $r$,
and constructs an estimate $\hx^k$ of $x^k$.

\textit{Distortion.}	
The average quadratic distortion 
between $x^k$ and $\hx^k$
is defined as
\begin{align}
\label{eq:Distortion}
    D \triangleq \frac{1}{k}\E{\norm{x^k - \hx^k}^{2}},
\end{align}	
where $\norm{\cdot}$ denotes the Euclidean norm, 
and the corresponding signal-to-distortion ratio (SDR)---by
\begin{align}
\label{eq:SDR}
    \SDR \triangleq \frac{\E{x_1^2}}{D}.
\end{align}	

\textit{Regime.}
We concentrate on the energy-limited regime, \viz\ the channel input is not subject to a bandwidth constraint, but rather to an energy constraint per \textit{source symbol} $E$~\eqref{eq:InputEnergyConstraint}.  
The per source-symbol capacity of the channel \eqref{eq:ChannelEq} is equal to \cite[Ch.~9.3]{CoverBook2Edition} 
\begin{align}
    \label{eq:ChannelCapacity_Lim}
    C = \ENR,
\end{align}
where $\ENR \triangleq E/N$ is the ENR,
and the capacity is measured in nats;
note that the available bandwidth is unconstrained (i.e., infinite).

Since the receiver can learn the noise level (for example by sacrificing some transmission time for training), we assume that the receiver has exact knowledge of the channel conditions. 
The transmitter is oblivious of the noise level, and needs to accommodate for a continuum of noise levels. Specifically, we will require the distortion to satisfy \eqref{eq:profile:tot}.
Throughout most of this work we will concentrate on the setting of infinite blocklength ($k \to \infty$). We will also conduct a simulation study for the scalar-source setting ($k=1$) in \secref{ss:numeric}.


\section{Background: Modulo-Lattice Modulation}
\label{s: MLM}

We will use MLM as a building block for robust JSCC with unknown ENR, where we will treat previous source estimators as effective side information (SI) known to the receiver but not to the transmitter \cite{JointWZ-WDP}, \cite[Ch.~11]{ZamirBook}.
We therefore review known results in this section for this technique and its application to Wyner--Ziv coding. 

We start by defining a sequence of seminorm-ergodic (SNE) vectors.

\begin{defn}[SNE {\cite[Def.~2]{OrdentlichErez_LatticeRobustness}}]
\label{def:SemiNorm}
    A sequence in $n$ of random vectors $z^{(n)}$ of length $n$ with a limit norm $\sigma_z > \infty$:\footnote{The original definition of \cite[Def.~2]{OrdentlichErez_LatticeRobustness} requires $\sigma^{(n)}_z = \sigma_z$ for all $n \in \nats$. We use here a more relaxed definition which will prove more convenient in the sequel.}
    \begin{align}
        \sigma^{(n)}_z &\triangleq \sqrt{ \frac{1}{n}\E{\norm{z^{({n})}}^2}}, &\lim_{n \to \infty} \sigma^{(n)}_z = \sigma_z ,
    \end{align}
    is SNE if for any $\eps, \delta > 0$, however small, there exists a large enough $n_0 \in \ints$, such that for all $n > n_0$
    \begin{align}
        \PR{ \frac{1}{n}\E{\norm{z^{(n)}}^2} >  (1+ \delta) \sigma_z^2 } \leq \eps .
    \end{align}
\end{defn}

We are now ready to present the model that will be considered in this section.

\textit{Source.} Consider a source sequence $x^k$ of length $k$, 
\begin{align}
    x^k = q^k + j^k ,
\end{align}
where $j^k$ is a SI sequence which is known to the receiver but not to the transmitter, and $q^k$ is the ``unknown part'' (at the receiver) 
with per-element variance 
\begin{align}
    \sigma_q^2 \triangleq \frac{1}{k} \E{\norm{q^k}^2}
\end{align}
and is 
SNE (as a sequence in $k$).

\textit{Transmitter.}
Maps $x^k$ to a channel input, $m^k$, that is subject to a power constraint
\begin{align}
    \frac{1}{k}\E{\norm{m^k}^2} \leq P.
\end{align}

\textit{Channel.}
The channel is an additive noise 
channel:
\begin{align}
\label{eq:MLM:output}
    y^k = m^k + z^k
\end{align}
where $z^k$ is an SNE noise vector that is 
uncorrelated with 
$x^k$ and has 
effective variance 
\begin{align}
    \sigma^2_z \triangleq \frac{1}{k}\E{\norm{z^k}^2}. 
\end{align}

The SNR is defined as $\SNR \triangleq P/\sigma^2_z$. 

\textit{Receiver.}
Receives $y^k$, in addition to the SI $j^k$, and generates an estimate $\hx^k \left(y^k, j^k\right)$ of the source $x^k$. 

The following MLM-based scheme will be employed in the sequel.

\begin{scheme}[MLM-based JSCC with SI {\cite{JointWZ-WDP}, \cite[Ch.~11]{ZamirBook}}]\ \\
\label{schm:MLM_Reznic}
    \textit{\textbf{Transmitter:}}
        Transmits the signal 
        \begin{align}
            \label{eq:MLM_Tx}
            m^k = [\eta x^k + d^k]_{\Lambda}
        \end{align}
        where $\Lambda$ is a lattice with a fundamental Voronoi cell $\mathcal{V}_0$ \cite[Ch.~2.2]{ZamirBook} and a second moment $P$ \cite[Ch.~3.2]{ZamirBook}, $\eta$ is a scalar scale factor, 
        $[\cdot]_\Lambda$ denotes the modulo-$\Lambda$ operation \cite[Ch.~2.3]{ZamirBook},
        and $d^k$ is a dither vector which is uniformly distributed over $\mathcal{V}_0$ and is independent of the source vector $x^k$; consequently, $m^k$ is independent of $x^k$ by the so-called crypto lemma \cite[Ch.~4.1]{ZamirBook}.
    
    \vspace{.5\baselineskip}
    \textit{\textbf{Receiver:}}
    
    \begin{itemize}
        \item
        Receives the signal $y^k$ \eqref{eq:MLM:output}
        and generates the signal 
        \begin{align}
        \label{eq:MLM_Rx1}
        \begin{aligned} 
            \ty^k &= [\alpha_c y^k - \eta j^k - d^k] _{\Lambda}
            \col{}{\\&}\triangleq [\eta q^k + z^k_\eff]_\Lambda
        \end{aligned} 
        \end{align}        
        where $z^k_\eff \triangleq -(1 - \alpha_c) m^k + \alpha_c z^k$ is the equivalent channel noise, and $\alpha_c$ is a channel scale factor.
        
        \item
        Generates an estimate $\hx^k$:
        \begin{align}
            \label{eq:MLM_Rx2}
            \hat{x}^k = \frac{\alpha_s}{\eta}\ty^k + j^k, 
        \end{align}         
        where $\alpha_s$ is a source scale factor.
    \end{itemize}
\end{scheme}

The following theorem provides guarantees for the achievable distortion using this scheme and is aggreagted 
from \cite{JointWZ-WDP}, \cite[Chs.~11.3, 6.4, 9.3]{ZamirBook}, and \cite{OrdentlichErez_LatticeRobustness} (see also the exposition about correlation-unbiased estimators (CUBEs) in \cite{RematchAndForward_Full}).

\begin{thm}
\label{thm:Reznic}
    The distortion \eqref{eq:Distortion} of \schemeref{schm:MLM_Reznic} is bounded from above by 
    \begin{align}
    \label{eq:MLM_Distortion}
        D &\leq L(\Lambda,P_e,\alpha_c)\cdot \tD + P_e\cdot D^\err ,
    \end{align} 
    for $\alpha_c \in (0,1], \alpha_s \in (0,1]$, and $\eta > 0$ that satisfy
    \begin{align}
        \frac{\eta^2\sigma^2_q}{P} + \frac{\alpha^2_c}{\SNR} + \left(1 - \alpha_c\right)^2 \leq 1,
    \end{align}
    where 
    \begin{align} 
        \label{eq:MLM_Dist_General}
        \tD \triangleq \left(1 - \alpha_s\right)^2\sigma^2_q + \alpha^2_s\left(\frac{\alpha^2_c}{\SNR} + \left(1 - \alpha_c\right)^2\right)\frac{P}{\eta^2} ,
    \end{align} 
    $D^\err$ is the distortion given a lattice decoding-error event \cite[Eq.~(24)]{JointWZ-WDP} and is bounded from above by 
    \begin{align}
    \label{eq:Derr}
        D^\err \leq 4 \sigma^2_q \left( 1 + \frac{\tL(\Lambda)}{\talpha} \right),
    \end{align}
    and the lattice parameters $L\left(\cdot,\cdot,\cdot\right)$ and $\tL(\cdot)$ are defined as 
    \begin{align}
    \label{eq:ChannelAndSelfNoise}
        L\left(\Lambda,P_e,\alpha_c\right) &\triangleq \mathrm{min}\left\{\ell: \Pr \left(\frac{z^k_\eff}{\sqrt{\ell}} \notin \mathcal{V}_0 \right) \leq P_e \right\} > 1 ,
     \\ \tL\left( \Lambda \right) &\triangleq \frac{\max_{a^k \in \mV_0} \norm{a^k}^2}{k P} > 1.
    \end{align}
    Moreover, for any $P_e > 0$, however small, and any $\alpha_c \in (0, 1]$, there exists a sequence of lattices, $\{ \Lambda_k | k \in \nats \}$, 
    that are good for both channel coding \cite[Def.~4]{OrdentlichErez_LatticeRobustness} and mean squared error (MSE) quantization \cite[Def.~5]{OrdentlichErez_LatticeRobustness}, viz.\ 
    \begin{align}
    \begin{aligned} 
        \lim_{k \to \infty} L(\Lambda_k, P_e, \alpha_c) &= 1
     \\ \lim_{k \to \infty} \tL(\Lambda_k) &= 1, 
    \end{aligned} 
    \label{eq:lattice-pars}
    \end{align}
    respectively, 
    and therefore this sequence of lattices achieves a distortion that approaches $\tD$.
\end{thm}
\begin{remark}
     By our definition of SNE sequences, for each finite $k$ the actual variance of the unknown part $\sigma^{(k)}_q$ and the noise variance $\sigma^{(k)}_z$ may be higher than for every $k < \infty$ their asymptotic quantities. 
     Consequently, also the second moment of  $\Lambda_k$ for every $k < \infty$ would be taken to be higher than its value asymptotic value.
 
     That said, as $k$ grows to infinity, these slacks become negligible and the performance converges to that of \eqref{eq:MLM_Distortion}, \eqref{eq:lattice-pars}.
\end{remark}

The following choice of parameters is optimal in the limit of infinite blocklength, $k \to \infty$, in the Gaussian case ($q^k$ comprises \iid\ Gaussian samples, $z^k$ comprises \iid\ Gaussian samples) \cite[Ch.~11.3]{ElGamalKimBook} when the SNR is known.

\begin{cor}[Optimal parameters {\cite{JointWZ-WDP}, \cite[Ch.~11.3]{ZamirBook}}]
\label{cor:MLM:opt-pars}
    The choice $\alpha_c = \alpha_c(\SNR)$, $L = L(\Lambda, P_e, \alpha_c)$, $\talpha = \talpha(\alpha_c, L)$, $\alpha_s(\SNR,\talpha, \alpha_c)$, $\eta = \eta(\talpha, \sigma_q^2)$
    yields a distortion $D$ that is bounded from above as in \eqref{eq:MLM_Distortion} with 

    \begin{align}
        \label{eq:MLM_Disotrtion_Limit}
        \tD &= \frac{\sigma^2_q}{1 + \talpha \cdot \left(1 + \SNR\right)},
    \end{align}
    where

    \begin{align}
        \alpha_c (\SNR) &\triangleq \frac{\SNR}{1 + \SNR}, 
     \\ \talpha(\alpha_c, L) &\triangleq \max\left(\alpha_c - \frac{L - 1}{L}, 0\right) \!,
     \\ \eta(\talpha, \sigma_q^2) &\triangleq \sqrt{ \talpha \frac{P}{\sigma^2_q} },
     \\ \alpha_s (\SNR, \talpha, \alpha_c) &\triangleq \frac{\SNR \cdot \talpha}{\SNR \cdot \talpha + \alpha_c}.
    \end{align}
    Moreover, for any $P_e > 0$, however small, there exists a sequence of lattices $\{ \Lambda_k | k \in \nats \}$ that attains \eqref{eq:lattice-pars} and therefore, in the limit $k \to \infty$,  $\talpha$ and $\alpha_s$ above converge to $\alpha_c$ and the distortion $D$ approaches $\tD$, which converges, in turn, to 
    \begin{align} 
        \tD = \frac{\sigma^2_q}{1 + \SNR} .
    \label{eq:OPTA}
    \end{align}
\end{cor}


Consider now the setting 
of an SNR that is unknown at the transmitter but is known at the receiver.\footnote{As discussed in \secref{s:Problem Statement}, we do not treat uncertainty at the receiver, as such uncertainty can be learned to any desired accuracy at negligibly cost.}
In this case, although 
the receiver knows the SNR and can therefore optimize $\alpha_c$ and $\alpha_s$ accordingly, the transmitter, being oblivious of the SNR, 
cannot optimize $\eta$ for the true value of the SNR. Instead, by setting $\eta$ in accordance with \corref{cor:MLM:opt-pars} for a preset minimal allowable design SNR, $\SNR_0$, 
\schemeref{schm:MLM_Reznic} achieves \eqref{eq:OPTA} for $\SNR = \SNR_0$ and improves, albeit sublinearly, with the SNR for $\SNR \geq \SNR_0$. This is detailed in the next corollary.

\begin{cor}[SNR universality]
\label{cor:SNR_Universal_MLM}
    Assume that $\SNR \geq \SNR_0$ for some predefined $\SNR_0 > 0$. 
    Then the choice $L(\Lambda, P_e, \alpha_c(\SNR_0))$, $\talpha = \talpha(\alpha_c(\SNR_0),L)$ and  $\eta = \eta(\talpha, \sigma_q^2)$ with respect to $\SNR_0$ (as it cannot depend on the true SNR), 
    and $\alpha_c = \alpha_c(\SNR)$ and $\alpha_s = \alpha_s(\SNR, \talpha, \alpha_c)$ (may depend on the true SNR) yields a distortion $D$ that is bounded from above as in \eqref{eq:MLM_Distortion} for $\tD$ that is given in  \eqref{eq:MLM_Disotrtion_Limit} with $\talpha = \talpha(\alpha_c(\SNR_0),L)$.
    Moreover, for any $P_e > 0$, however small, there exists a sequence of lattices $\{ \Lambda_k | k \in \nats \}$ that satisfies \eqref{eq:lattice-pars}; therefore, in the limit $k \to \infty$,  $\talpha$ converges to $\alpha_c(\SNR_0)$, $\alpha_s$---to $\frac{\SNR_0(1 + \SNR)}{\SNR_0 (1 + \SNR) + 1 + \SNR_0}$, and the distortion $D$ approaches $\tD$ which converges, in turn, to 
    \begin{align}
        \tD &= \frac{\sigma^2_q}{1 + \SNR}\frac{1}{\frac{1}{1 + \SNR} + \frac{\SNR_0}{1 + \SNR_0}}.
    \end{align} 

\end{cor}

\begin{cor}[Source-power uncertainty]
\label{rem:Unknown_SI_Power}
    Assume now additionally that the transmitter is oblivious of the exact power of $q^k$, $\sigma^2_q$, but knows that it is bounded from above by $\tsigma^2_q$: $\sigma^2_q \leq \tsigma^2_q$.
    Then 
    the distortion is bounded according to \eqref{eq:MLM_Distortion}~with 
    \begin{align}
        \label{eq:MLM_Disotrtion_Universal_UnknownSI}
        \tD &= \frac{\tsigma^2_q}{\frac{\tsigma^2_q}{\sigma^2_q} + \talpha \cdot \left(1 + \SNR\right)}
    \end{align}
    for the parameters 
    \begin{align}
    \label{eq:OptimalPrms_UnknownSrc}
    \begin{aligned} 
        \alpha_c &= \frac{\SNR}{1 + \SNR}, 
     \col{\quad}{\\} \talpha &= \talpha(\alpha_c(\SNR_0),L),
     \col{\quad}{\\} \eta &= \eta(\talpha, \tsigma_q^2),
     \col{\quad}{\\}  \alpha_s &= \frac{\talpha\left(1 + \SNR\right)}{\frac{\tsigma^2_q}{\sigma^2_q} + \talpha\left(1 + \SNR\right)}, 
    \end{aligned} 
    \end{align}
    Moreover, for any $P_e > 0$, however small, there exists a sequence of lattices $\{ \Lambda_k | k \in \nats \}$ that attains \eqref{eq:lattice-pars} and therefore, in the limit of $k \to \infty$,  $\talpha$ converges to $\alpha_c(\SNR_0)$, $\alpha_s$---to $\frac{1 + \SNR}{(1+\SNR) + \frac{\tsigma^2_q}{\sigma^2_q}\frac{1 + \SNR_0}{\SNR_0}}$, and the distortion $D$ is bounded from above in this limit by $\tD$: 
    \begin{subequations}
    \begin{align}
        \label{eq:MLM_Disotrtion_Universal_step_UnknownSource}
        D &\leq \tD + \eps 
    \\ &= \frac{\tsigma^2_q}{1 + \SNR}\cdot\frac{1}{\frac{\tsigma^2_q}{\sigma^2_q}\cdot\frac{1}{1 + \SNR} + \frac{\SNR_0}{1 + \SNR_0}} + \eps 
     \\ &\leq \min \left\{ \frac{\sigma_q^2}{1 + \SNR_0}, \frac{\tsigma_q^2}{1 + \SNR} \frac{1 + \SNR_0}{\SNR_0} \right\} + \eps ,
    \label{eq:MLM_Distortion_BoundFinal}
    \end{align}
    where $\eps$ decays to zero with $P_e$.
    \end{subequations}
    For $\SNR \geq \SNR_0 \gg 1$, the 
    bound \eqref{eq:MLM_Distortion_BoundFinal} approaches $\frac{\tsigma^2_q}{1 + \SNR}$.
\end{cor}

The following result is a simple consequence of \thmref{thm:Reznic} and avoids exact computation of the optimal parameters.
\begin{cor}[Suboptimal parameters]
\label{cor:MLM:correlated_params}
    Assume the setting of \corref{rem:Unknown_SI_Power} but with $z^k$
    not necessarily uncorrelated with $m^k$, 
    and denote $\SDR = P / \sigma_z^2$.\footnote{We refer to it by $\SDR$ since now $z^k$ may depend on $m^k$.}
    Then, the distortion is bounded according to \eqref{eq:MLM_Distortion} with 
    \begin{align}
        \tD &= \frac{\tsigma^2_q}{\SDR}
    \end{align}
    for the parameters 
    $\talpha = \alpha_c = \alpha_s = 1$, $\eta = \eta(1, \tsigma_q^2)$.
\end{cor} 

The following property will prove useful in \secref{s: UnKnownENR_SchemeDesign}.
\begin{lem}[{\cite[Lemmata 6 and 11]{ErezZamirAWGN}}] 
\label{lem:f_dither<f_G}
    Let $\{\Lambda_k | k \in \nats \}$ be a sequence of lattices that satisfies the results in this section, and let $d^k$ be a dither that is uniformly distributed over the fundamental Voronoi cell of $\Lambda_k$. Then, the probability density function (p.d.f.) of $d^k$ is bounded from above as 
        \begin{align} 
            f_{d^k}(a^k) &\leq f_{G^k}(a^k) \e^{\eps_k k} &\forall a^k \in \reals^k ,
        \end{align} 
        where $f_{G^k}$ is the p.d.f.\ of a vector with i.i.d.\ Gaussian entries with zero mean and the same second moment $P$ as $\Lambda_k$, and $\eps_k > 0$ decays to zero with $k$. 
\end{lem}


\section{Background: Analog Modulations in the Known-ENR Regime}
\label{s: AnalogPPM}

In this section, we review analog modulations for conveying a scalar zero-mean Gaussian source ($k=1$) over a channel with infinite bandwidth, where both the receiver and the transmitter know the channel noise level, or equivalently, $\ENR = E / N$.

Consider first analog linear modulation, in which the source sample $x$ is linearly transmitted with energy $E$,\footnote{Under linear transmission, the energy constraint holds only on average, and the transmit energy is equal to the square of the specific realization of $x$.} using some 
unit-energy waveform
\begin{align}
\label{eq:Background_KnownENR_DirectLinear}
    s_{x}(t) = \sqrt{E} \frac{x}{\sigma_x} \varphi(t).
\end{align} 
Note that linear modulation is the same (``universal'') regardless of the true noise level.
Signal space theory
\cite[Ch.~8.1]{WozencraftJacobsBook},  \cite[Ch.~2]{ViterbiOmuraBook}
suggests that 
a sufficient statistic of the transmission of \eqref{eq:Background_KnownENR_DirectLinear} over the channel \eqref{eq:ChannelEq} is 
the one-dimensional projection $y$ of $r$ onto $\varphi$: 
\begin{align}
    \label{eq:Background_KnownENR_DirectLinearScalar}
\begin{aligned} 
    y &= \int_{-\frac{T}{2}}^{\frac{T}{2}} \varphi(t) r(t) dt
 \col{}{\\ &}= \sqrt{E}\frac{x}{\sigma_x} + \sqrt{\frac{N}{2}} z ,
\end{aligned} 
\end{align}
where $z$ is a standard Gaussian noise variable.
The minimum mean square error (MMSE) estimator of $x$ from $y$ is linear and its distortion is equal to 
\begin{align}
\label{eq:Background_KnownENR_DirectLinearMSE}
    D &= \frac{\sigma^2_x}{1 + 2\ENR},
\end{align}
and improves only linearly with the $\ENR$.

Consider now analog PPM, in which the source sample is modulated by the shift of a given pulse rather than by its amplitude (which is the case for analog linear modulation):
\begin{align}
\label{eq:PPM:waveform}
    s_x(t) = \sqrt{E}\phi(t - x \Delta)
\end{align}
where $\phi$ is a predefined pulse with unit energy and $\Delta$ is a scaling parameter.
In particular, the square pulse,\footnote{Clearly, the bandwidth of this pulse is infinite. By taking a large enough bandwidth $W$, one may approximate this pulse to an arbitrarily high precision and attain its performance within an arbitrarily small gap.} is known to achieve good performance. 
This pulse is given by 
\begin{align}
    \label{eq:AnalogPPM:PulseShaping_RectPulse}
    \phi(t) &=
    \begin{cases}
        \sqrt{\frac{\beta}{\Delta}}, & \abs{t} \leq \frac{\Delta}{2\beta},\\
        0, & \mathrm{otherwise},
    \end{cases}
\end{align}    
for a parameter $\beta > 1$ which is sometimes referred to as \textit{effective dimensionality}. Clearly, $T = \Delta + \Delta/\beta$. 

The optimal receiver is the MMSE estimator $\hx$ of $x$ given the entire output signal:
\begin{align}
    \hx^\MMSE = \CE{x}{r}.
\label{eq:delay-estimate:MMSE}
\end{align}
The following theorem provides an upper bound on the achievable distortion of this scheme using  (suboptimal) maximum a posteriori (MAP) decoding, which is given by 
\begin{align}
\label{eq:AnalogPPM:MAP:Gaussian}
    \hx^\MAP &= \argmax_{a\in\reals} \left\{ R_{r, \phi}(a \Delta) - \frac{N}{4\sqrt{E}}a^2 \right\},
\end{align}
where 
\begin{subequations} 
\label{eq:correlations}
\noeqref{eq:Rphi}
\begin{align} 
\begin{aligned} 
    R_{r,\phi}(\hx \Delta) &\triangleq \int_{-\infty}^\infty r(t)\phi(t - \hx \Delta) dt
 \col{}{\\ &}= \sqrt{E} R_\phi \left( (x-\hx) \Delta \right) + \sqrt{\frac{\beta}{\Delta}} \int_{\hx \Delta - \frac{\Delta}{2\beta}}^{\hx \Delta + \frac{\Delta}{2\beta}} n(t) dt ,
\end{aligned}
\label{eq:Rr_phi}
\end{align} 
is 
the (empirical) cross-correlation function between 
$r$ and $\phi$ with lag (displacement)~$\hx \Delta$,
and 
\begin{align}
\begin{aligned} 
    R_{\phi} (\tau) &= \int_{-\infty}^\infty \phi(t) \phi(t - \tau) dt 
 \col{}{\\ &}= 
    \begin{cases}
        1 - \frac{|\tau|}{\frac{\Delta}{\beta}}, & |\tau| \leq \frac{\Delta}{\beta}
     \\ 0, & \mathrm{otherwise}
    \end{cases}
\end{aligned}
\label{eq:Rphi}
\end{align}
\end{subequations}
is the autocorrelation function of $\phi$ with lag $\tau$.

\begin{remark}
    Since a Gaussian source has infinite support, the required overall transmission time $T$ is infinite. Of course this is not possible in practice. Instead, one may limit the transmission time $T$ to a very large---yet finite---value. This will incur a loss compared to the the bound that will be stated next; this loss can be made arbitrarily small by taking $T$ to be large enough.
\end{remark}

\begin{thm}[\!\!{\cite[Prop.~2]{EnergyLimitedJSCC:Lev_Khina:Full}}]
\label{thm:UpperBound_GaussianPrior}
    The distortion of the MAP decoder \eqref{eq:AnalogPPM:MAP:Gaussian} of a standard Gaussian scalar source transmitted using analog PPM with a rectangular pulse is bounded from above by 
    \begin{align}
        D \leq D_S + D_L 
    \end{align}
    with
    \begin{align}
        &D_L \triangleq 2\beta\sqrt{\ENR}\e^{-\frac{\ENR}{2}}\Bigg(1 + 3\sqrt{\frac{2\pi}{\ENR}} + \frac{12\e^{-1}}{\beta\sqrt{\ENR}} + \frac{8\e^{-1}}{\sqrt{8\pi}\beta} 
    \nonumber
    \\* &\ + \sqrt{\frac{8}{\pi\ENR}} +  \frac{12^{\frac{3}{2}}\e^{-\frac{3}{2}}}{\beta\sqrt{32\pi\ENR}}\Bigg)
        + \beta\sqrt{8\pi}\e^{-\ENR}\left(1 + \frac{4\e^{-1}}{\beta\sqrt{2\pi}}\right),
    \nonumber
     \\ &D_S \triangleq \frac{\frac{13}{8} + \sqrt{\frac{2}{\beta}}\left(\sqrt{2 \beta \ENR} - 1 \right) \cdot \e^{-\left(\sqrt{\ENR} - \frac{1}{\sqrt{2 \beta}}\right)^2} }{\left(\sqrt{\beta \ENR} - \frac{1}{\sqrt{2}} \right)^4} 
      + \frac{\e^{-\beta\ENR}}{\beta^2} ,
    \nonumber
    \end{align}
    bounding the small- and large-error distortions, 
    assuming $\beta \ENR > 1/2$.
    In particular, in the limit of large $\ENR$, and $\beta$ that increases monotonically with $\ENR$, 
    \begin{align}
     \label{eq:UpperBounf_GaussianPrior_Explicit_HighENR}
         D \leq \left(\tD_S + \tD_L \right)\{1 + o(1)\}
    \end{align}
    where  
    \begin{align}
         \tD_S &\triangleq\frac{13/8}{\left(\beta\ENR\right)^2},\\
         \tD_L &\triangleq 2\beta\sqrt{\ENR}\cdot\e^{-\frac{\ENR}{2}} ,
    \end{align} 
    and $o(1) \to 0$ in the limit of $\ENR \to \infty$.
\end{thm}

\begin{remark}
\label{rem:AnaolgPPM:quadratic-improvement}
    For a fixed $\beta$, the distortion improves quadratically with the $\ENR$. This behavior will proof useful in the next section, where we construct schemes for the unknown-ENR regime. 
\end{remark}

Setting $\beta = \left(\frac{13}{8}\right)^{\frac{1}{3}}\left(\ENR\right)^{-\frac{5}{6}} \e^{\frac{\ENR}{6}}$ in \eqref{eq:UpperBounf_GaussianPrior_Explicit_HighENR} of \thmref{thm:UpperBound_GaussianPrior} yields the following asymptotic performance.
    
\begin{cor}[\!\!{\cite[Th.~2]{EnergyLimitedJSCC:Lev_Khina:Full}}]
\label{thm:analog-PPM:knownENR:Gaussian:achievable}
    The achievable distortion of a standard Gaussian scalar source transmitted over an energy-limited channel with a known ENR is bounded from above as 
    \begin{align}
    \label{eq:KnownENR:RectPulse_UpperBoundOptim_Final:GaussianSource}
        D &\leq 3\cdot\left(\frac{13}{8}\right)^{\frac{1}{3}}\, \e^{-\frac{\ENR}{3}}\cdot \left(\ENR\right)^{-\frac{1}{3}} \cdot \left\{1 + o(1)\right\}
        ,
    \end{align}
    where 
    $o(1) \to 0$ as $\ENR \to \infty$.
\end{cor} 

The following corollary, whose proof is available in the appendix, states that the (bound on the) distortion is continuous in the source p.d.f.\ around a Gaussian p.d.f.
Such continuity results of the MMSE estimator in the source p.d.f.\ are known \cite{Wu-Verdu:MMSE:IT2011}. 
Next, we prove the required continuity directly for our case of interest with an additional technical requirement on the deviation from a Gaussian p.d.f.;
this result will be used in conjunction with a non-uniform variant of the Berry--Esseen theorem in \secref{s: UnKnownENR_SchemeDesign}.
\begin{cor}
\label{cor:NearGaussianDist_UpperBound}
    Consider the setting of \thmref{thm:UpperBound_GaussianPrior} for a source p.d.f.\ that satisfies 
    \begin{align}
    \label{eq:pdf-deviation}
        \abs{f_x(a) - f_G(a)} &\leq \eps \delta_f(a) , & \forall a &\in \reals ,
    \end{align}
    where $\eps > 0$; $f_G$ is the standard Gaussian p.d.f.; and $\delta_f$ is a symmetric absolutely-continuous non-negative bounded function  
    with unit integral, $\int_{\infty}^\infty \delta_f(a) da = 1$, that is 
    monotonically decreasing for $x > 0$ (and for $x < 0$, by symmetry) and satisfies $\delta_f(x) \in o\left(x^{-4}\right)$; thus, there exists $H < \infty$ such that 
    \begin{align} 
    \label{eq:delta_f=o(x^4)}
        \delta_f(x) &\leq \frac{H}{(1 + x)^4}, &\forall x \in \reals.
    \end{align} 
    
    Then, the distortion of the decoder that applies the decoding rule \eqref{eq:AnalogPPM:MAP:Gaussian}
    is bounded from above by\footnote{\label{foot:MAP-dec} This is no longer the MAP decoding rule since $f_x$ is no longer a Gaussian p.d.f.} 
    \begin{align}
        D \leq D_G + \eps C ,
    \end{align}
    where $D_G = D_S + P_L D_L$ denotes the bound on the distortion for a standard Gaussian source of \thmref{thm:UpperBound_GaussianPrior}, and $C < \infty$ is a non-negative constant that depends on $\delta_f$.
\end{cor}

	
\section{Main Results}
\label{s: UnKnownENR_SchemeDesign}

In this section, we construct JSCC solutions for the unknown-ENR regime communications problem.
Since an exponential improvement with the ENR cannot be attained in this setting \cite{MinimumEnergyBound_Tuncel}, following \cite{baniasadi2020minimum,MinimumEnergyBound_Tuncel}, 
we consider polynomially decaying profiles \eqref{eq:Profile_Polynom}.

We construct an MLM-based layered scheme where each layer accommodates a different noise level, with layers of lower noise levels acting as SI in the decoding of subsequent layers.

We first show in \secref{s:LinearTransmissions_MLM} that replacing the successive refinement coding of \cite{baniasadi2020minimum,MinimumEnergyBound_Tuncel} with MLM (Wyner--Ziv coding) with \textit{linear} layers results in better performance in the infinite-bandwidth setting (paralleling the results of the bandwidth-limited setting \cite{Reznic}).

In \secref{s:SquaredProfile}, we replace the last layer with an analog PPM one, which improves quadratically with the ENR [$L=2$ in \eqref{eq:Profile_Polynom}] above the design ENR (recall \remref{rem:AnaolgPPM:quadratic-improvement}).

In principle, despite analog PPM attaining a gracious quadratic decay with the ENR (recall \remref{rem:AnaolgPPM:quadratic-improvement}) only above a predefined design ENR, since the distortion is bounded from above by the (finite) variance of the source, 
it attains a quadratic decay with the ENR for all $\ENR \in \reals$, or equivalently, 
for all $N \in \reals$ and $L=2$ in \eqref{eq:Profile_Polynom}.

That said, the performance of analog PPM deteriorates rapidly when the ENR is below the design ENR of the scheme, meaning that the minimum energy required to obtain \eqref{eq:profile:tot} with $L=2$ and a given $\Edsgn$ is large. To alleviate this, we use the above-mentioned layered MLM scheme. 
Furthermore, to achieve higher-order improvement with the ENR [$L>2$ in \eqref{eq:Profile_Polynom}], multiple layers in the MLM scheme need to be employed. 

We compare the analytic and empirical results of the proposed scheme in \secref{ss:numeric}.

We now present a simplified variant of the general scheme that is considered throughout this section. This variant is also depicted in \figref{fig:MLM_JSCC_Scheme_Simpler}. 
The full scheme, which incorporates interleaving for analytical purposes, is available in \appref{app:Full_M_Layer} and depicted in \figref{fig:MLM_JSCC_Scheme}. 

\begin{scheme}[MLM-based]\ \\
 \label{schm:UniversalAnalogMLM}
    \textit{\textbf{$M$-Layer Transmitter:}}
    
    \textit{First layer ($i=1$):}
    \begin{itemize}
    \item 
        Transmits each of the entries of the vector $x^k$ over the channel \eqref{eq:ChannelEq} linearly \eqref{eq:Background_KnownENR_DirectLinear}:
        \begin{align}
        \nonumber
            s_{1;\ell}(t) &\triangleq s\left(t+\left(\ell-1\right) T\right) = \sqrt{\frac{E_1}{T}} \frac{x_\ell}{\sigma_x} \varphi(t), 
            & t \in [0,T),
        \end{align}
        for $\ell \in \{1, \ldots, k\}$, 
        where $\varphi$ 
        is a 
        continuous 
        unit-norm (i.e., unit-energy) 
        waveform that is zero outside the interval $[0,T]$, say $\phi$ of \eqref{eq:AnalogPPM:PulseShaping_RectPulse}, 
        $E_1 \in [0, E]$ is the allocated energy for layer $1$, and $E$ is the total available energy of the scheme.
    \end{itemize}

    \textit{Other layers:} For each $i \in \{2, \ldots, M\}$:
        \begin{itemize}
        \item 
            Calculates the $k$-dimensional tuple 
            \begin{align}
            \label{eq:RepeatedQuant_Tx}
                m^k_i &= [\eta_i x^k + d^k_i]_{\Lambda_i} \,, 
            \end{align}
            where $m^k_i = \begin{bmatrix} m_{i;1} & m_{i;2} & \ldots & m_{i;k} \end{bmatrix}^\dagger$, and $m_{i;\ell}$ denotes the $\ell^\mathrm{th}$ entry of $m^k_i$; 
            $\eta_i$, $d^k_i$ and $\Lambda_i$ take the roles of $\eta,d^k$ and $\Lambda$ of \schemeref{schm:MLM_Reznic}, and are tailored for each layer $i$; $\Lambda_i$ is chosen to have unit second moment.
        \item 
            For each $\ell \in \{1, \ldots, k\}$, 
            views $m_{i;\ell}$ 
            as a scalar source sample, 
             and generates a corresponding channel input, 
            \begin{align} 
                s_{i;\ell}(t) &\triangleq s\left(t + (\ell - 1) T + (i - 1) k T\right), & t \in [0,T),
            \end{align} 
            using a scalar JSCC scheme with a predefined energy $E_i \geq 0$ that is designed for a predetermined $\ENR_i$, or equivalently, $N_{i} = E_i/\ENR_i$, such that $\sum_{i=1}^{M} E_i = E$ and $N_2 > N_3 > \cdots > N_{M} > 0$. 
            
         \end{itemize} 

\vspace{.5\baselineskip}
    \textit{\textbf{Receiver:}} Receives the channel output signal $r$ \eqref{eq:ChannelEq}, and recovers the different layers as follows.

    \textit{First layer ($i=1$):}
        For each $\ell \in \{1, \ldots, k\}$:
        \begin{itemize}
        \item 
            Recovers the MMSE estimate $\hx_{1;\ell}$ of $x_\ell$ given 
            $\{ r_{1;\ell}(t) | t \in [0, T) \}$, 
            where $r_{1;\ell}(t) \triangleq r(t + (\ell-1) T)$.
        \item 
            If the true noise level $N$ satisfies $N > N_2$, sets the final estimate $\hx_\ell$ of $x_\ell$ to $\hx_{1;\ell}$ 
            and stops. Otherwise, determines the maximal layer index $\jmath \in \{2, \ldots, M\}$ for which $N \leq N_\jmath$ and continues to process the other layers. 
        \end{itemize}
        
    \textit{Other layers:}
        For each $i \in \{ 2, \ldots, \jmath \}$ in ascending order:
        \begin{itemize}
        \item 
            For each $\ell \in \{1, \ldots, k\}$, uses the receiver of the scalar JSCC scheme to generate an estimate 
            $\htm_{i;\ell}$ of $\tm_{i;\ell}$
            from 
            $\left\{ r_{i;\ell}(t) |  t \in [0, T) \right\}$, 
            where
            \begin{align}
                r_{i;\ell}(t) \triangleq r\left(t + (\ell-1) T + (i-1) k T \right) .
            \end{align}
        \item
            Using the effective channel output $\hm_i^k$ (that takes the role of $y^k$ in \schemeref{schm:MLM_Reznic}) with SI $\hx_{i-1}^k$, generates the signal 
            \begin{align}
                \label{eq:MLM:UnknownENR:Layers}
                \ty^k_i &= [\alpha^{(i)}_c \hm^k_i - \eta_i \hx^k_{i-1} - d^k_i]_{\Lambda_i},
            \end{align}
            as in \eqref{eq:MLM_Rx1} of \schemeref{schm:MLM_Reznic}, 
            where $\alpha^{(i)}_c$ is a channel scale factor.
        \item 
            Constructs an estimate $\hx_i^k$ of $x^k$:
            \begin{align}
                \hx_i^k &= \frac{\alpha^{(i)}_s}{\eta_i}\ty^k_i + \hx^k_{i-1},
            \end{align}
            as in \eqref{eq:MLM_Rx2} of \schemeref{schm:MLM_Reznic}, 
            where $\alpha^{(i)}_s$ is a source scale factor. The final estimate if $\hx^k = \hx^k_\jmath$.
     \end{itemize}
 \end{scheme} 
    
\begin{remark}[Interleaving]
\label{rem:Interleaving}
    To guarantee independence between all the noise entries $\ell \in \{1, \ldots, k\}$, we use interleaving in the full scheme, which is described in \appref{app:Full_M_Layer} in \eqref{eq:interleave} and \eqref{eq:deinterleave}.
    We note that this operation is used to simplify the proof that the resulting noise vector is SNE (recall \defnref{def:SemiNorm}). 
\end{remark} 

\begin{remark}[Gaussianization]
\label{rem:Gaussianization}
    To use the analysis of \secref{s: AnalogPPM} of analog PPM for a Gaussian source, 
    we multiply the vectors $m^k_i$ by orthogonal matrices $\mH_i$ that effectively ``Gaussianize'' its entries, as shown in the full description of the scheme in \appref{app:Full_M_Layer}, in \eqref{eq:interleave} and \eqref{eq:deinterleave}. 
    In particular, this is achieved by a Walsh--Hadamard matrix $\mH_i$ by appealing to the central limit theorem;
    a similar choice was previously proposed by Feder and Ingber \cite{Feder-Ingber:Patent:Hadamard2012}, and by 
    Hadad and Erez 
    \cite{HadadErez:Dithered:Gaussianize:SP2016}, where in the latter, 
    the columns of the Walsh--Hadamard matrix were further multiplied by \iid\ Rademacher RVs to achieve near-independence between multiple descriptions of the same source vector
    (see \cite{Asnani-Shomorony-Avestimehr-Weissman:WorstCaseCompression:Gaussianize:IT2015,NoWeissman:RDF:Extremes:IT2016,HadadErez:Dithered:Gaussianize:SP2016} for other ensembles of orthogonal matrices that achieve a similar result).
    Interestingly, the multiplication by the orthogonal matrices $\mH_i^{-1} = \mH_i^\dagger$ (since Walsh--Hadamard matrices are symmetric, they further satisfy $\mH_i^\dagger = \mH_i$) Gaussianizes the effective noise incurred at the outputs of the analog PPM JSCC receivers.
\end{remark}

\begin{remark}[JSCC-induced channel]
\label{rem:induced-channel}
    The continuous-time JSCC transmitter and receiver over the infinite-bandwidth AWGN channel induce an effective additive-noise channel of better effective SNR and source's bandwidth. Over this induced channel, the MLM transmitter and receiver are then employed. 
    This interpretation is depicted in \figref{fig:JSCC_Effective_channel} with $\tn_i^k$ representing the effective additive noise vectors.
\end{remark}

\begin{figure*}[!]
    \begin{subfigure}[t]{\textwidth}
    \centering
	\col{\resizebox{.7\textwidth}{!}{\input{figures/MLM_JSCC_Full_Toli3_OneCol.tikz}}}
	{\resizebox{.73\textwidth}{!}{\input{figures/MLM_JSCC_Full_Toli3.tikz}}}
	\caption{Full scheme}
    \label{fig:MLM_JSCC_Scheme_Simpler}
    \end{subfigure}
    \begin{subfigure}[t]{\textwidth}
    \centering
	\resizebox{0.81\textwidth}{!}{\pgfmathsetseed{4}

\begin{tikzpicture}[auto, arrow/.style={very thick, ->, >=stealth'},node distance=5mm,>=latex']
    \node [coord] (input) {};
    \node [coord, below = 5mm of input] (input_for_frame) {};
    \node [coord, right = 166mm of input] (output_for_frame) {};

    
    \node[block, below = 6mm of input_for_frame, fill=none, draw=none] (MLM_Tx_0) {};
    \node[block, right = 35mm of MLM_Tx_0, align = center] (MLM_Tx_1) {MLM Tx$_2$};
    \node[block, fill=none, draw=none, right = 20mm of MLM_Tx_1, align = center, minimum width = 12mm] (MLM_Tx_dots) {$\bullets$};
    \node[block, right = 20mm of MLM_Tx_dots, align = center] (MLM_Tx_last) {MLM Tx$_{M}$};
    \node[coord, below = 4mm of MLM_Tx_0] (JSCC_Tx_0_midmidarrow) {};    
    
    \node[sum, below = 11mm of MLM_Tx_0] (channel_JSCC0) {\bf \Large +};
    \node[coord, left of = channel_JSCC0, node distance = 10mm] (noise0) {};
    \node[coord, left of = channel_JSCC0, node distance = 15mm] (noise0_for_frame) {};
    \node[coord, below = 6mm of channel_JSCC0] (channel_output0) {};
    \node[coord, below = 6mm of MLM_Tx_0] (channel_JSCC0_for_frame) {};
    
    \node[sum, below = 11mm of MLM_Tx_1] (channel_JSCC1) {\bf \Large +};
    \node[coord, left of = channel_JSCC1, node distance = 10mm] (noise1) {};
    \node[coord, left of = channel_JSCC1, node distance = 15mm] (noise1_for_frame) {};
    \node[coord, below = 6mm of channel_JSCC1] (channel_output1) {};
     \node[coord, below = 6mm of MLM_Tx_1] (channel_JSCC1_for_frame) {};
     
    \node[sum, below = 11mm of MLM_Tx_last] (channel_JSCC_last) {\bf \Large +};
    \node[coord, left of = channel_JSCC_last, node distance = 10mm] (noise_last) {};
    \node[coord, left of = channel_JSCC_last, node distance = 16mm] (noise_last_for_frame) {};
    \node[coord, below = 6mm of channel_JSCC_last] (channel_output_last) {};
    \node[coord, below = 6mm of MLM_Tx_last] (channel_JSCC_last_for_frame) {};
    
    \node[block, below = 5mm of channel_output0, fill=none, draw=none] (MLM_Rx_0) {
    };
    \node[block, right = 35mm of MLM_Rx_0, align = center] (MLM_Rx_1) {MLM Rx$_2$};
    \node[block, fill=none, draw=none, right = 20mm of MLM_Rx_1, align = center, minimum width = 12mm] (MLM_Rx_dots) {$\bullets$};
    \node[block, right = 20mm of MLM_Rx_dots, align = center] (MLM_Rx_last) {MLM Rx$_{M}$};
    
    \node[block, below = 9mm of MLM_Rx_0, fill=none, draw=none] (output_0) {
    };
    \node[block, below = 9mm of MLM_Rx_1, fill=none, draw=none] (output_1) {
    };
    \node[block, below = 9mm of MLM_Rx_dots, fill=none, draw=none] (output_dots) {
    };
    \node[block, below = 9mm of MLM_Rx_last, fill=none, draw=none] (output_last) {
    };
    \node[coord, below = 5mm of MLM_Rx_0, fill=none, draw=none] (SI_0) {
    };
    \node[coord, below = 5mm of MLM_Rx_1, fill=none, draw=none] (SI_1) {
    };
    \node[coord, below right = 5mm and 2mm of MLM_Rx_dots, fill=none, draw=none] (SI_last) {
    };
    
    \draw[arrow] (input) -| node[right, pos=.0] {$x^k$} (channel_JSCC0);
    \draw[arrow] (input_for_frame) -| node[below]         {}      (MLM_Tx_1);
    \draw[arrow,dotted] (input_for_frame) -| node[below]  {}      (MLM_Tx_dots);
    \draw[arrow] (input_for_frame) -| node[below]         {}      (MLM_Tx_last);

    \draw[arrow, very thick](MLM_Tx_1) -- node[right, pos=.35] {$m_2^k$} (channel_JSCC1);
    \draw[arrow, very thick](MLM_Tx_last) -- node[right, pos=.35] {$m^k_{M}$} (channel_JSCC_last);

    \draw[arrow] (noise0) -- node[left, pos=0.1] {$\tn^{k}_1$} (channel_JSCC0);
    \draw[arrow] (channel_JSCC0) -- node [right, pos=.97] {$\hx^{k}_1$} (output_0);
    \draw[arrow] (noise1) -- node[left, pos=0.1] {$\tn^{k}_2$} (channel_JSCC1);
    \draw[arrow] (channel_JSCC1) -- node {$\hm^{k}_2$} (MLM_Rx_1);
    \draw[arrow] (MLM_Rx_1) -- node [pos = .89] {$\hx^{k}_2$} (output_1);
    \draw[arrow] (noise_last) -- node[left, pos=0.1] {$\tn^{k}_M$} (channel_JSCC_last);
    \draw[arrow] (channel_JSCC_last) -- node {$\hm^{k}_M$} (MLM_Rx_last);
    \draw[arrow] (MLM_Rx_last) -- node [pos = .89] {$\hx^{k}_M$} (output_last);
    
    \draw[arrow] (SI_0) -| node[below, pos=.03] {} ($(MLM_Rx_1.south west) + (2.5mm,0)$);;
    \draw[arrow, dotted] (SI_1) -| node[below, pos=.03] {} ($(MLM_Rx_dots.south west) + (2.5mm,0)$);
    \draw[arrow, dotted] (SI_last) -| node[above, pos=.25] {$\hx^{k}_{M - 1}$} ($(MLM_Rx_last.south west) + (2.5mm,0)$);
    
    \begin{pgfonlayer}{background}
        \node[draw, rounded corners, dashed, fit = (channel_JSCC0) (noise0_for_frame)] (jscc0) {};
    \end{pgfonlayer}
    \node[right = 0 of jscc0] {$\begin{array}{l} \text{Effective} \\ \text{channel 1} \end{array}$}; 

    \begin{pgfonlayer}{background}
        \node[draw, rounded corners, dashed, fit = (channel_JSCC1) (noise1_for_frame)] (jscc1) {};
    \end{pgfonlayer}
    \node[right = 0 of jscc1] {$\begin{array}{l} \text{Effective} \\ \text{channel 2} \end{array}$}; 
    
        \begin{pgfonlayer}{background}
        \node[draw, rounded corners, dashed, fit = (channel_JSCC_last) (noise_last_for_frame)] (jscc_last) {};
    \end{pgfonlayer}
    \node[right = 0 of jscc_last] {$\begin{array}{l} \text{Effective} \\ \text{channel } M \end{array}$}; 
    
\end{tikzpicture}
    }
	\vspace{-1.75\baselineskip}
	\caption{With effective additive-noise channel}
    \label{fig:JSCC_Effective_channel}
    \end{subfigure}
    \caption{Block diagrams of \schemeref{schm:UniversalAnalogMLM} and of this scheme with the effective additive-noise channels of \remref{rem:induced-channel}.}
\end{figure*} 

We next provide analytic guarantees for this scheme, for linear and analog PPM layers in \secref{s:LinearTransmissions_MLM} and \secref{s:SquaredProfile}, respectively, 
in the infinite-blocklength regime. 
In \secref{ss:numeric}, we compare the analytic and empirical performance of these schemes in the infinite-blocklength regime, as well as 
compare the empirical performance of these schemes for a single source sample.
The treatment of the infinite-blocklength regime pertains to the full scheme as presented in \appref{app:Full_M_Layer}. 
The comparison for  
a single source sample,  
uses the simplified variant of \schemeref{schm:UniversalAnalogMLM}.

\subsection{Infinite-Blocklength Setting with Linear Layers}
\label{s:LinearTransmissions_MLM}

We start with analyzing the performance of 
the scheme
where all the $M$ layers are transmitted linearly and $M$ is large; 
we concentrate on the setting of an infinite source blocklength ($k \to \infty$) and derive an achievability bound on the minimum energy that achieves a distortion profile \eqref{eq:Profile_Polynom}. 
The following theorem is proved in \appref{app:proof:thm:linear_MLM}.

\begin{thm} 
\label{thm:linear_MLM}
    Choose a decaying order $L > 1$, a design parameter $\Edsgn > 0$, and a minimal noise level $\Nmin > 0$, however small.
    Then, a distortion profile \eqref{eq:profile:tot} with $L$ and $\Edsgn$ is achievable for all noise levels $N > \Nmin$ for any transmit energy $E$ that satisfies 
    \begin{align}
        \label{eq:MinimumEnergy_L2}
        E > \delta_{\mathrm{lin}}\left(L\right) \Edsgn
        ,
    \end{align}
    for a large enough source blocklength $k$, 
    where 
    \begin{align}
        \delta_{\mathrm{lin}}\left(L\right) &\triangleq \frac{1}{2}\cdot\min_{(\alpha, x) \in \reals^2_{+} }\Bigg\{\left(\frac{\e^{\alpha}}{x}\right)^{L - 1} 
    \col{}{\\* &\ \ } + \frac{x}{2}\left(\e^{\alpha L} - 1\right)\left(1 + \sqrt{1 + \frac{4\e^{\alpha\left(L + 1\right)}}{\left(1 - \e^{\alpha L}\right)^2}}\right)\frac{\e^{-2\alpha}}{1 - \e^{-\alpha}}\Bigg\}.
    \end{align}
    In particular, the choice $\left( x, \alpha \right) = (0.898,0.666)$ achieves a quadratic decay ($L=2$) for any transmit energy $E$ that satisfies 
    \begin{align}
    \label{eq:MinimalEnergy_Linear_quad}
        E > 2.167 \Edsgn,
    \end{align}
    for a large enough source blocklength $k$.
\end{thm}

We note that already this variant of the scheme offers an improvement compared to the hitherto best known upper (achievability) bound of \eqref{eq:minimum-energy-bounds}.

The choice of the minimal noise level $\Nmin$ dictates the number of layers $M$ that need to be employed: The lower $\Nmin$ is, the more layers $M$ need to be employed.

\begin{remark}
\label{rem:ExponentialDecayAllocation}
    In the proof in \appref{app:proof:thm:linear_MLM}, we use an exponentially-decaying noise-level series: $N_i = \Delta \e^{-\alpha \left(i - 1\right)}$, which facilitates the analysis. Nevertheless, any other assignment that satisfies the profile requirement and energy constraint is valid and may lead to better performance; for further discussion, see \secref{s:Summary}.
\end{remark}

\subsection{Infinite-Blocklength Setting with Analog PPM Layers}
\label{s:SquaredProfile}

In this section, we concentrate on the setting of an infinite source blocklength ($k \to \infty$) and a quadratically decaying profile [$L=2$ in \eqref{eq:profile:tot}] using analog PPM.

To that end, we use a sequence of $M-1$ linear JSCC layers as in \secref{s:LinearTransmissions_MLM}, with only the last layer replaced by an analog PPM one; since analog PPM improves quadratically with the ENR (recall \remref{rem:AnaolgPPM:quadratic-improvement}), $M$ need not go to infinity to attain a quadratically decaying profile. 

\begin{thm} 
\label{thm:quadratic_profile}
    Choose a design parameter $\Edsgn > 0$, and a minimal noise level $\Nmin > 0$, however small.
    Then, a quadratic profile ($L=2$) \eqref{eq:Profile_Polynom} with $\Edsgn$ is achievable for all noise levels $N > \Nmin$ for any transmit energy $E$ that satisfies
    \begin{align}
        E > 1.961 \Edsgn ,
    \label{eq:Emin:analog_ppm_mlm}
    \end{align}
    for a large enough source blocklength $k$.
\end{thm}

This theorem, whose proof is available in \appref{app:proof:thm:quadratic_profile}, 
offers a further improvement over the upper bounds in \eqref{eq:minimum-energy-bounds} and \thmref{thm:linear_MLM} for a quadratic profile. 

\begin{remark} 
    Replacing all layers, but the first layer, with analog PPM ones should yield better performance,  
    but complicates the analysis. 
    Moreover, similar analysis to that of \thmref{thm:linear_MLM} for $L \neq 2$ may be devised, but for $L > 2$ would require multiple layers as the distortion of analog PPM decays only quadratically.
    Both of these analyses are left for future research.
\end{remark} 


\section{Simulations}
\label{ss:numeric}

\col{
\begin{figure}[t]
	\begin{subfigure}[t]{.5\columnwidth}
		\centering
	    \includegraphics[width=\columnwidth]{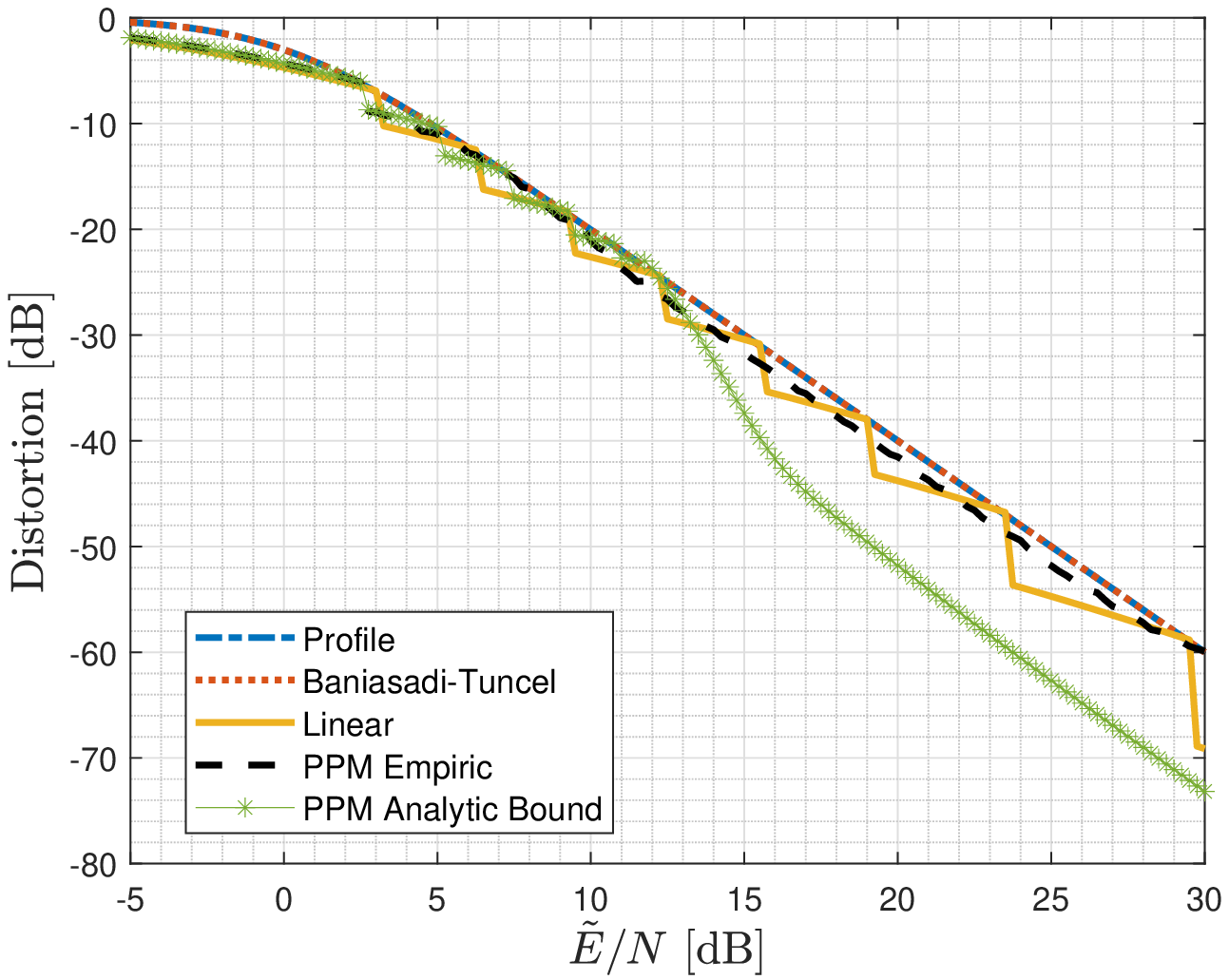}
	\end{subfigure}
	\begin{subfigure}[t]{.5\columnwidth}
		\centering
	    \includegraphics[width=\columnwidth]{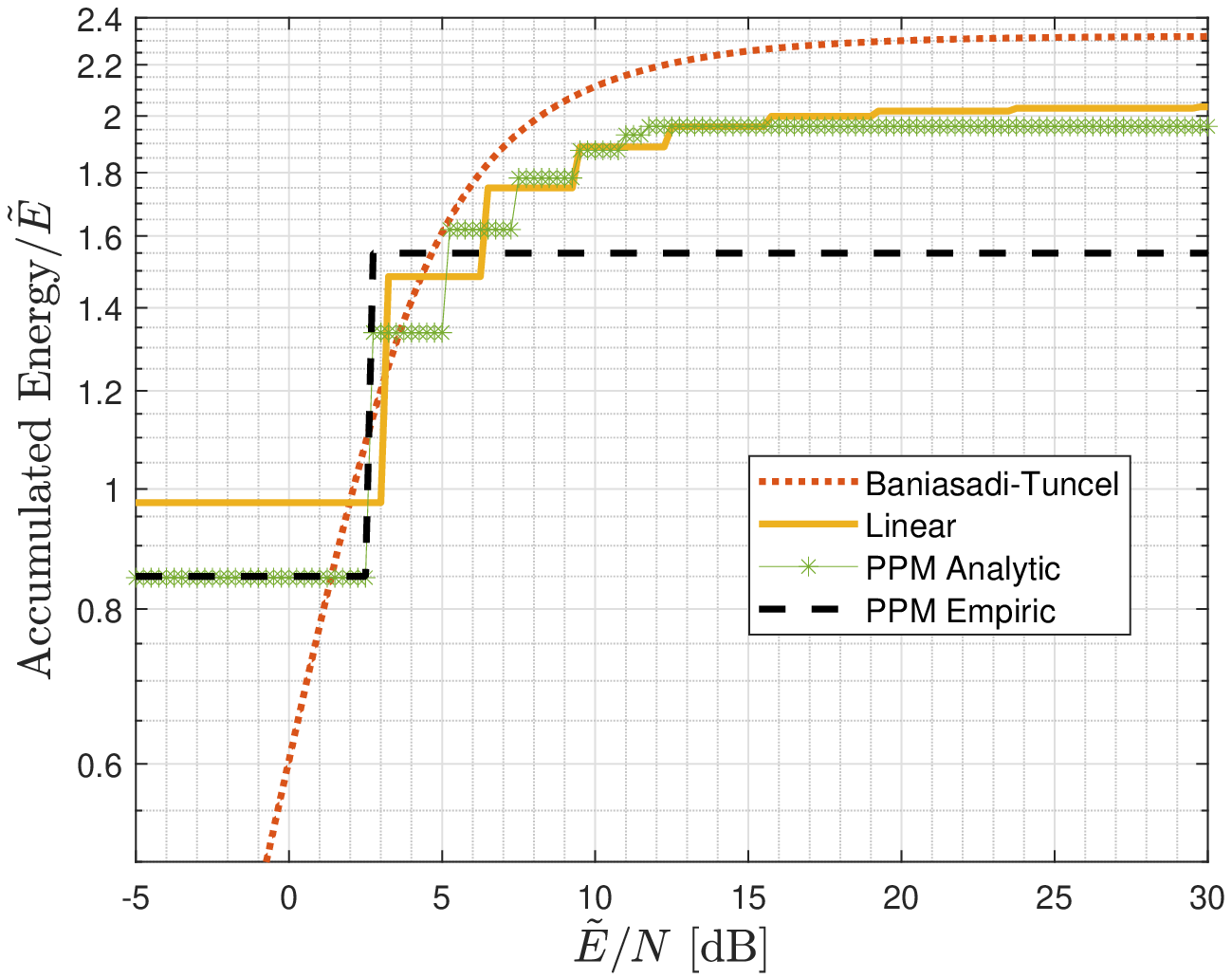}
	\end{subfigure}
	\caption{Distortion and accumulated energy of the layers utilized by the receiver at a given $\Edsgn/N$ for a Gaussian source in the infinite-blocklength regime for a quadratic profile: 
\schemeref{schm:UniversalAnalogMLM} with linear layers with energy allocation $E_i = \Delta \e^{-\alpha i}$ for  $\Delta = 0.975, \alpha = 0.65$, empirical performance of the scheme with a linear layer with energy $E_1 = 0.85$ and an analog PPM layer with energy $E_2 = 0.75$, and analytic performance of the scheme 
of \thmref{thm:quadratic_profile} with the parameters from its proof and analytic performance of Baniasadi and Tuncel scheme according to the proof in \cite{baniasadi2020minimum}}
	\label{fig:UnknownENR_InfiniteDim}
\end{figure}
}{
\begin{figure}[t]
		\vspace{-.8\baselineskip}
	\begin{subfigure}[t]{\columnwidth}
		\centering
	    \includegraphics[width=\columnwidth]{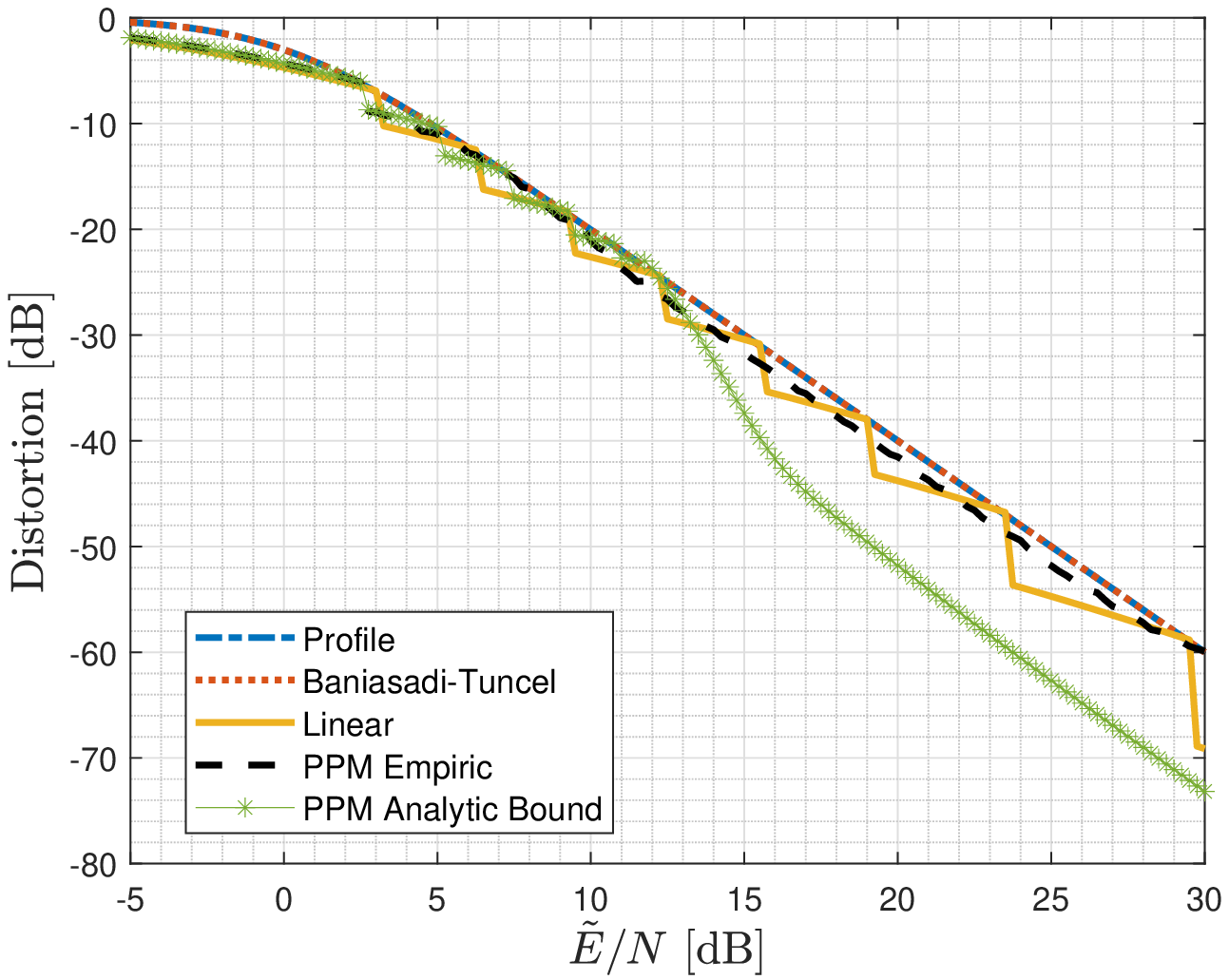}
	\end{subfigure}
	\begin{subfigure}[t]{\columnwidth}
		\centering
	    \includegraphics[width=\columnwidth]{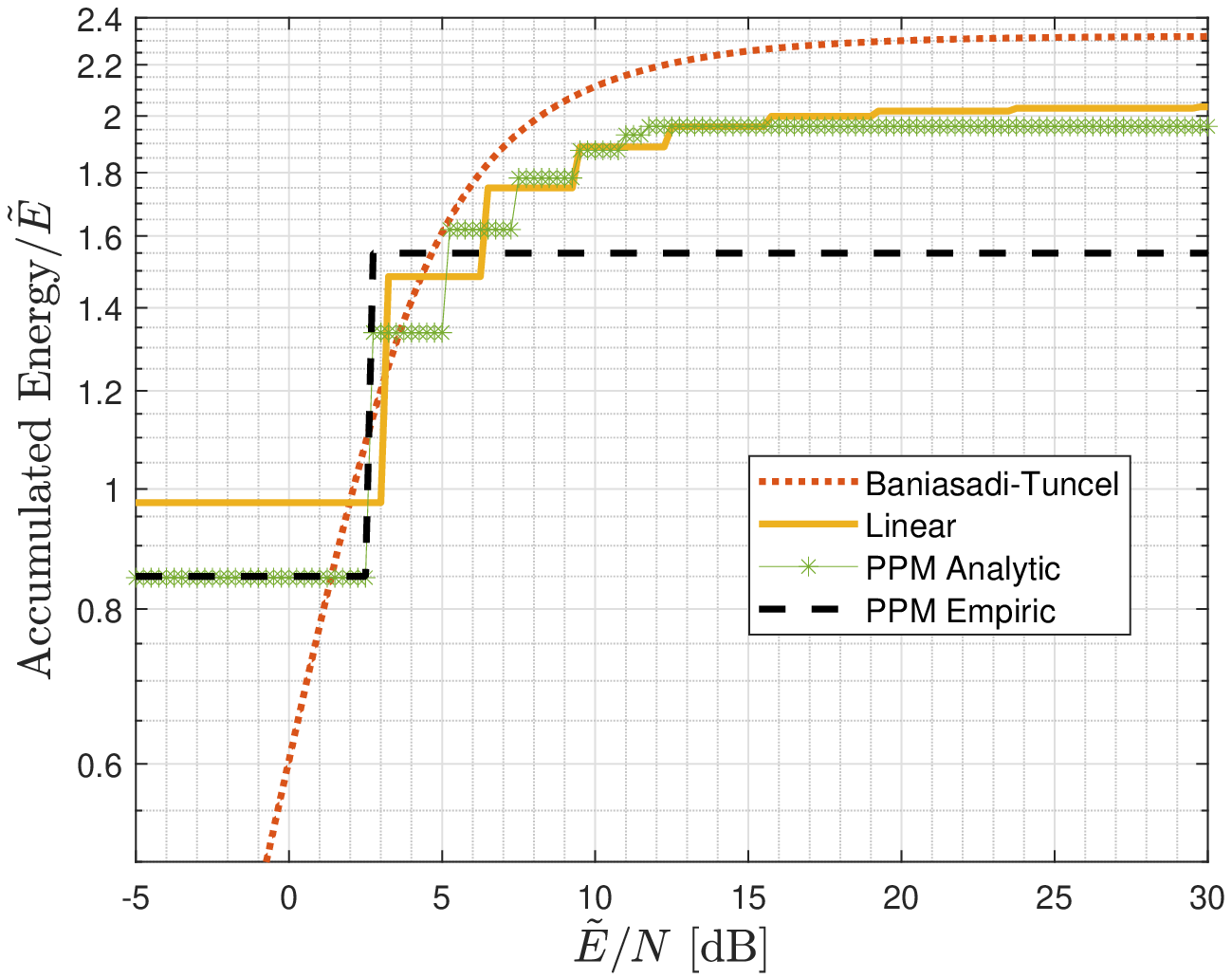}
	\end{subfigure}
	\caption{Distortion and accumulated energy of the layers utilized by the receiver at a given $\Edsgn/N$ for a Gaussian source in the infinite-blocklength regime for a quadratic profile: 
\schemeref{schm:UniversalAnalogMLM} with linear layers with energy allocation $E_i = \Delta \e^{-\alpha i}$ for  $\Delta = 0.975, \alpha = 0.65$, empirical performance of the scheme with a linear layer with energy $E_1 = 0.85$ and an analog PPM layer with energy $E_2 = 0.75$, analytic performance of the scheme 
of \thmref{thm:quadratic_profile} with the parameters from its proof and analytic performance of Baniasadi and Tuncel scheme according to the proof in \cite{baniasadi2020minimum}}
	\label{fig:UnknownENR_InfiniteDim}
\end{figure}
}

\col{
\begin{figure}[t]
	\begin{subfigure}[t]{.5\columnwidth}
		\centering
	    \includegraphics[width=\columnwidth]{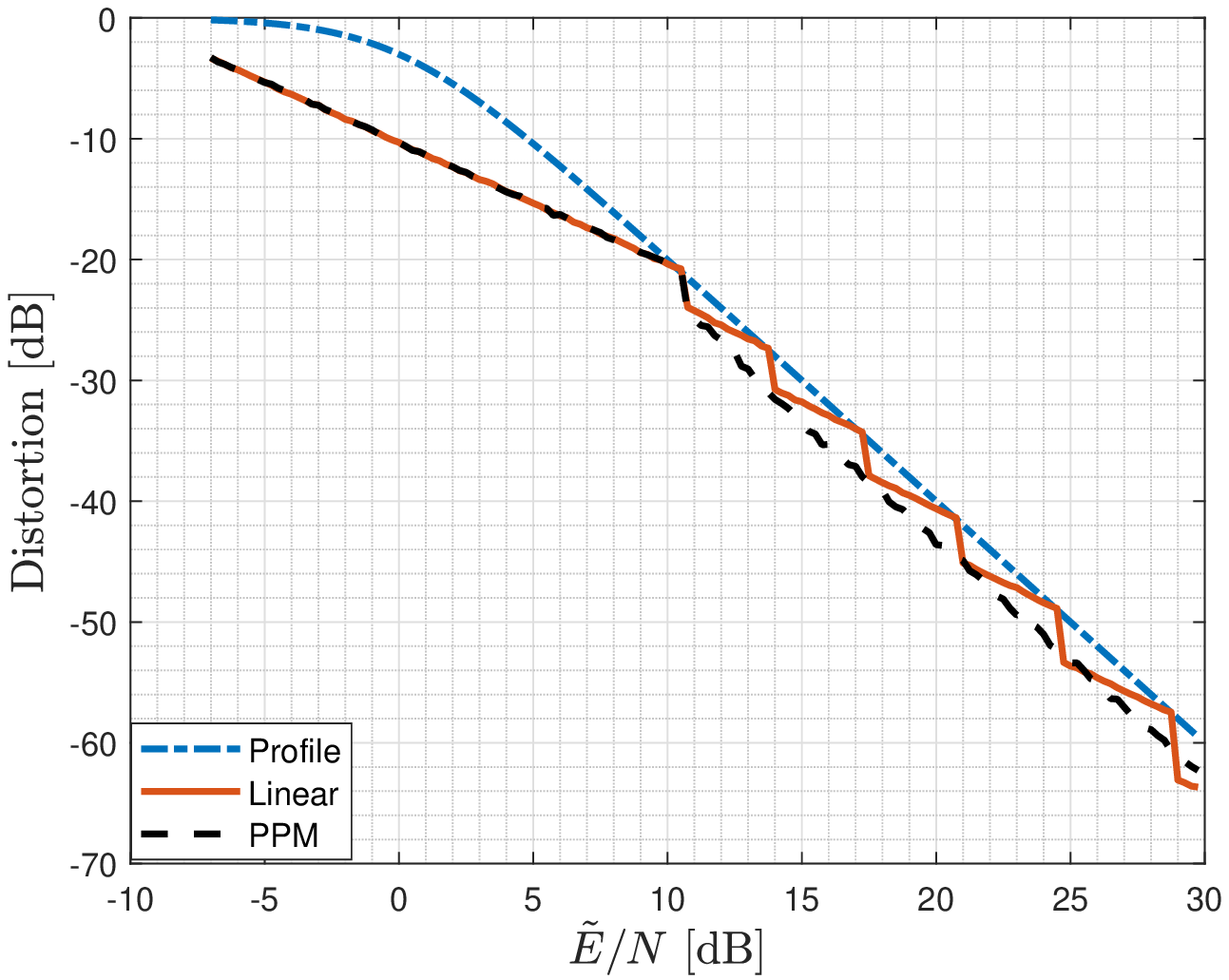}
	\end{subfigure}
	\begin{subfigure}[t]{.5\columnwidth}
		\centering
	    \includegraphics[width=\columnwidth]{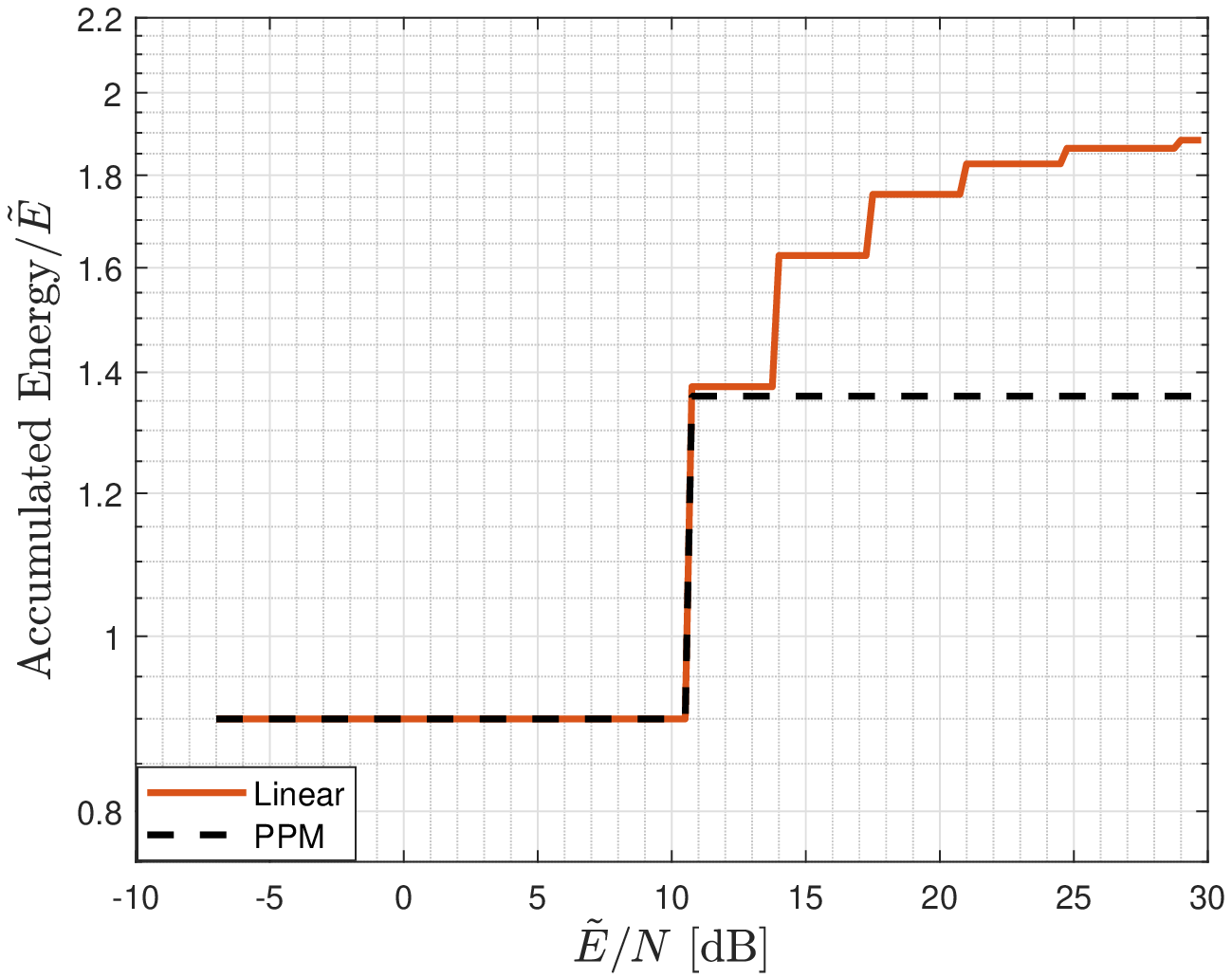}
	\end{subfigure}
	\caption{Distortion and accumulated energy of the layers utilized by the receiver at a given $\Edsgn/N$ for a uniform scalar source for a quadratic profile: 
\schemeref{schm:UniversalAnalogMLM} with linear layers with energy allocation $\frac{E_i}{\Edsgn} = \Delta \e^{-\alpha i}$ for  $\Delta = 0.9, \alpha = 0.64$, and with a linear layer with energy $E_1 = 0.9\Edsgn$ and an analog PPM layer with energy $E_2 = 0.346\Edsgn$.}
	\label{fig:UnknownENR_Scalar}
\end{figure}
}{
\begin{figure}[t]
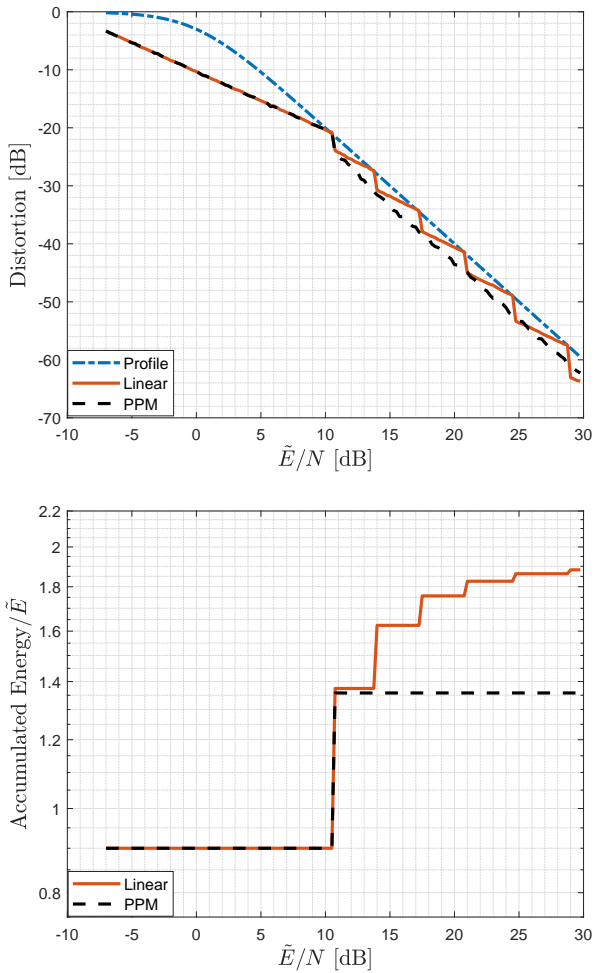

		\vspace{-.8\baselineskip}
	\begin{subfigure}[t]{\columnwidth}
		\centering
	    \includegraphics[width=\columnwidth]{figures/TotalFigure_MLM_PPM_Scalar_Distortion.eps}
	\end{subfigure}
	\begin{subfigure}[t]{\columnwidth}
		\centering
	    \includegraphics[width=\columnwidth]{figures/TotalFigure_MLM_PPM_Scalar_Energy.eps}
	\end{subfigure}
	\caption{Distortion and accumulated energy of the layers utilized by the receiver at a given $\Edsgn/N$ for a uniform scalar source for a quadratic profile: 
\schemeref{schm:UniversalAnalogMLM} with linear layers with energy allocation $\frac{E_i}{\Edsgn} = \Delta \e^{-\alpha i}$ for  $\Delta = 0.9, \alpha = 0.64$, and with a linear layer with energy $E_1 = 0.9\Edsgn$ and an analog PPM layer with energy $E_2 = 0.346\Edsgn$.}
	\label{fig:UnknownENR_Scalar}
\end{figure}
}

We first consider the infinite-blocklength regime ($k \to \infty$) for a Gaussian source and a quadratic profile [$L=2$ in \eqref{eq:profile:tot}], for which we have derived analytical guarantees in Secs.~\ref{s:LinearTransmissions_MLM} and \ref{s:SquaredProfile}.
\figref{fig:UnknownENR_InfiniteDim} depicts the accumulated energy of the employed layers at the receiver of \schemeref{s: UnKnownENR_SchemeDesign}  
and the achievable distortion at a given $\Edsgn/N$, along with the desired quadratic distortion profile \eqref{eq:Profile_Polynom} (with $L=2$) for $\Nmin \to 0$
for:
linear layers,
and
$M-1$ linear layers with a final analog PPM layer (analytic performance for $M=7$ layers according to \thmref{thm:quadratic_profile} and empirical performance for $M=2$ layers).
This figure clearly demonstrates the gain due to introducing an analog PPM layer. Interestingly, the empirical curve shows that only two layers are needed when the second layer is an analog PPM one, meaning that the seven layers needed in the proof of \thmref{thm:quadratic_profile} are an artifact of the slack in our analytic bounds.
To derive the performance of the scheme with linear layers we evaluated \eqref{eq:proof:unknown-ENR:Di} directly for  
the optimized energy allocation $E_i = \Delta \e^{-\alpha i}$ with  $\Delta = 0.975$ and $\alpha = 0.65$. 
To derive the analytical performance of \thmref{thm:quadratic_profile}, we used the energy allocation from its proof in \appref{app:proof:thm:quadratic_profile}, 
while for the empirical performance, optimizing over the energy allocation yielded $E_1 = 0.975 \Edsgn, E_2 = 0.5904 \Edsgn$. 

We move now to the uniform scalar source setting ($k=1$) and a quadratic profile. 
The analysis of \schemeref{s: UnKnownENR_SchemeDesign} in the scalar setting is difficult. We therefore evaluate its performance empirically for both variants of the scheme: with linear layers, and with one linear layer and one analog PPM layer (two layers suffice in this setting as well).
In \figref{fig:UnknownENR_Scalar}, we depict again the accumulated energy of the employed layers at the receiver of \schemeref{s: UnKnownENR_SchemeDesign} 
and the achievable distortion at a given $\Edsgn/N$ for both variants of the scheme, along with the desired quadratic distortion profile \eqref{eq:Profile_Polynom} (with $L=2$) for $\Nmin \to 0$.


\section{Summary and Discussion}
\label{s:Summary}

In this work, we studied the problem of JSCC over an energy-limited channel with unlimited bandwidth and/or transmission time when the noise level is unknown at the transmitter. We showed that MLM-based schemes outperform the existing schemes thanks to the improvement in the performance of all layers (including preceding layers that act as SI) with the ENR. By replacing (some of the) linear layers with analog PPM ones, further improvement was achieved. 
We further demonstrated numerically that the MLM-layered scheme works well in the scalar-source regime.

We also note that a substantial gap remains between the lower bound in \eqref{eq:minimum-energy-bounds} and the upper bound of \thmref{thm:quadratic_profile} 
for the energy required to achieve a quadratic profile [\eqref{eq:Profile_Polynom} with $L=2$].
In \secref{s:FutureResearch} several ways to close this gap are described.

We note that, although we assumed that both the bandwidth and the time are unlimited, the scheme and analysis presented in this work carry over to the setting where one of the two is bounded as long as the other one is unlimited, with little adjustment.

\section{Future Research}
\label{s:FutureResearch}

Consider first the remaining gap between the lower and upper bounds. As demonstrated in \secref{ss:numeric}, the upper (achievability) bound on the performance of analog PPM is not tight and calls for further improvement thereof. 
This step is currently under intense investigation, along with improvement via companding of the presented analog PPM variant in this work as well as via other choices of energy allocation (see \remref{rem:ExponentialDecayAllocation}). 
Furthermore, the optimization was performed numerically and for a particular form of noise levels of an exponential form (recall \remref{rem:ExponentialDecayAllocation}). We believe that a systematic optimization procedure could put light on the weaknesses of our scheme and provide further improvement of the overall performance. 
On the other hand, the outer bounds of \cite{baniasadi2020minimum} are based on specific choices of sequences of noise levels. Therefore, further improvement might be achieved by other choices and calls for further research. 

We have also shown that the MLM scheme performs well in the scalar-source regime; it would be interesting to derive analytical performance guarantees for this regime.

Finally, since MLM utilizes well source SI at the receiver and channel SI at the transmitter \cite{JointWZ-WDP,AnalogMatching}, \cite[Chs.~10--12]{ZamirBook},
the proposed scheme can be extended to limited-energy settings such as universal transmission with universal SI at the receiver
\cite{baniasadi-tuncel-SI:ISIT2021} and the dual problem of the one considered in this work of universal transmission with near-zero bandwidth~\cite{baniasadi-tuncel:ZeroBW:ISIT2020}.


\appendices


\section{Proof of \corref{cor:NearGaussianDist_UpperBound}}
\label{app:proof:cor:NearGaussianDist_UpperBound}

To prove \corref{cor:NearGaussianDist_UpperBound}, we repeat the steps of the proof of \thmref{thm:UpperBound_GaussianPrior} in \cite[Prop.~2]{EnergyLimitedJSCC:Lev_Khina:Full};
we next detail the contributions to the small-distortion \cite[Eq.~(25)]{EnergyLimitedJSCC:Lev_Khina:Full} and the large-distortion \cite[Eq.~(27)]{EnergyLimitedJSCC:Lev_Khina:Full} terms due to the deviation \eqref{eq:pdf-deviation} from the source p.d.f.\ from Gaussianity, which are denoted by $d_S$ and $d_L$, respectively.

We start by bounding the contribution to the small-distortion term.
To that end, note that \cite[Eqs.~(24b) and (25b)]{EnergyLimitedJSCC:Lev_Khina:Full} remain unaltered since the decoder remains the same. 
The contribution to the small-distortion term is bounded from above as follows.
\begin{subequations}
\label{eq:deviation:small-distortion}
\begin{align}
     \frac{d_{S}}{\eps} &\leq \frac{2}{\beta^2}\cdot\int_{\sqrt{2\beta\ENR}}^{\infty}\delta_f(a) da
\label{eq:deviation:small-distortion:delta}
 \\ &\leq \frac{2}{\beta^2}
\label{eq:deviation:small-distortion:h-pdf}
\end{align}
\end{subequations}
where \eqref{eq:deviation:small-distortion:delta} follows from \cite[Eqs.~(24b) and (25b)]{EnergyLimitedJSCC:Lev_Khina:Full},
and \eqref{eq:deviation:small-distortion:h-pdf} follows from $\delta_f$ being non-negative with unit integral.

We next bound the contribution $d_L$ to the large distortion term.
To that end, note that \cite[Eqs.~(27) and (28)]{EnergyLimitedJSCC:Lev_Khina:Full} remain unaltered since the decoder remains the same. 
We define by $a_i$ the deviation in $\PR{A_i}$ in \cite[Eq.~(30)]{EnergyLimitedJSCC:Lev_Khina:Full}. Then,
\begin{subequations}
\label{eq:eps_L}
\begin{align}
    \frac{a_i}{\eps} 
    &\leq \int_{-\left(\frac{2\ENR\beta}{i} + \frac{i}{2\beta}\right)}^{\infty} \delta_f (a) \Bigg\{ \frac{\sqrt{3}}{4\pi}\e^{-\frac{\ell^2(a)}{3}} \! + \left(\frac{1}{\sqrt{8}} + \frac{\ell(a)}{4\sqrt{\pi}}\right)\e^{-\frac{\ell^2(a)}{4}}
\col{}{\nonumber
\\* &\quad} + \e^{-\frac{\ell^2(a)}{2}} \Bigg\}da \col{\nonumber
\\* &\quad}{} + \int_{-\infty}^{-\left(\frac{2\ENR\beta}{i} + \frac{i}{2\beta}\right)} \delta_f(a) da 
\label{eq:eps_L_step1}
 \\ &\leq \frac{\sqrt{2\ENR}\beta}{i}\int_{0}^{\infty}\delta_f\left(\frac{\sqrt{2\ENR}\beta}{i}u - \frac{2\ENR\beta}{i} - \frac{i}{2\beta}\right)
\col{}{\nonumber
\\* &\ \qquad\quad} \!\Bigg\{ \frac{\sqrt{3}}{4\pi}\e^{-\frac{u^2}{3}} + \e^{-\frac{u^2}{2}} + \left(\frac{1}{\sqrt{8}} + \frac{u}{4\sqrt{\pi}}\right)\e^{-\frac{u^2}{4}} \Bigg\}du
    \nonumber
    \\* & \qquad\quad + \frac{H}{\left(1 + \frac{2\ENR\beta}{i} + \frac{i}{2\beta}\right)^4}
\label{eq:eps_L_step2}
 \\ &\leq \frac{\tH}{\left(1 + \frac{2\ENR\beta}{i} + \frac{i}{2\beta}\right)^4} \,,
\label{eq:eps_L_step3}
\end{align}
\end{subequations}
where 
\eqref{eq:eps_L_step1} follows from \cite[Eqs.~(28) and (30)]{EnergyLimitedJSCC:Lev_Khina:Full},
\eqref{eq:eps_L_step2} follows from integration by substitution and \eqref{eq:delta_f=o(x^4)}, 
and \eqref{eq:eps_L_step3} follows from \eqref{eq:delta_f=o(x^4)} for some $\tH > 0$.

By substituting the bound of \eqref{eq:eps_L} in \cite[Eq.~(31)]{EnergyLimitedJSCC:Lev_Khina:Full}, we may bound $d_L$ from above by 
\begin{align} 
\label{eq:d_L}
    \frac{d_L}{\eps} &\leq 2 \sum_{i=2}^\infty \left( \frac{i}{\beta} \right)^2 a_i
 \col{}{\\ &}\leq \sum_{i=2}^\infty \left( \frac{i}{\beta} \right)^2 \frac{\tH}{\left(1 + \frac{2\ENR\beta}{i} + \frac{i}{2\beta}\right)^4}
 \col{}{\\ &}\leq \tC 
\end{align} 
for some $\tC < \infty$.

Therefore, by \eqref{eq:deviation:small-distortion} and \eqref{eq:d_L}, 
the overall contribution $d$ to the distortion due to the deviation \eqref{eq:pdf-deviation} is bounded from above by 
\begin{align}
    d &= d_L + d_S 
 \col{}{\\ &}\leq \eps \left( \frac{2}{\beta^2} + \tC \right);
\end{align}
choosing $C = \frac{2}{\beta^2} + \tC < \infty$ concludes the proof.


\section{Full version of \schemeref{schm:UniversalAnalogMLM}} 
\label{app:Full_M_Layer}

We now present the full multi-layer transmission scheme (cf.~\schemeref{schm:UniversalAnalogMLM}), which includes interleaving and Gaussianization steps, as discussed in Rems.~\ref{rem:Interleaving} and \ref{rem:Gaussianization}, respectively. 
Block diagrams of the overall scheme and the new ingredients are provided in Figs.~\ref{fig:MLM_JSCC_Scheme} and \ref{fig:SubBlocks_PiTxRx}, respectively.  
The new components in  \schemeref{schm:UniversalAnalogMLM_Full} compared to those in \schemeref{schm:UniversalAnalogMLM} (and \figref{fig:MLM_JSCC_Scheme_Simpler}) are highlighted in green in \figref{fig:MLM_JSCC_Scheme}. 

\col{
\begin{figure*}[t]
    \centering
	{\center
	\resizebox{0.8\textwidth}{!}{\input{figures/MLM_JSCC_Full_WithInterleaver_updated_OneCol.tikz}}}
	\caption{Block diagram of \schemeref{schm:UniversalAnalogMLM_Full}.}
	\vspace{-0.75\baselineskip}
    \label{fig:MLM_JSCC_Scheme}
\end{figure*}   
}
{
\begin{figure*}[th]
    \centering
	{\center
	\resizebox{.8\textwidth}{!}{\input{figures/MLM_JSCC_Full_WithInterleaver_updated.tikz}}}
	\caption{Block diagram of \schemeref{schm:UniversalAnalogMLM_Full}.}
	\vspace{-0.75\baselineskip}
    \label{fig:MLM_JSCC_Scheme}
\end{figure*}
}
\col{
\begin{figure}
\begin{subfigure}[t]{0.5\columnwidth}
	\centering
    \resizebox{\textwidth}{!}{\pgfmathsetseed{4}

\begin{tikzpicture}[auto, arrow/.style={very thick, ->, >=stealth'},node distance=5mm,>=latex']
    \node [coord] (input) {};
    \node [coord, right = 25mm of input] (input_for_frame) {};

    
    \node[coord, right = 6mm of input] (Base_MLM_node) {};
    \node[block, above right = 10mm and 8mm of Base_MLM_node, align = center] (MLM_Tx_0) {MLM \\ Tx$_i$};
    \node[block, below = 5mm of MLM_Tx_0, fill=none, draw=none, align = center] (MLM_Tx_dots) {$\bullets$};
    \node[block, below = 5mm of MLM_Tx_dots, align = center] (MLM_Tx_last) {MLM \\ Tx$_i$};
    \node[coord, right = 5mm of Base_MLM_node, fill=none, draw=none, align = center] (MLM_Mid_Block) {};
    \draw[arrow] (input) -- node[above, pos=.03] {$\mX$} (MLM_Mid_Block);
    \draw[arrow,thin] (MLM_Mid_Block) |- node[left, pos=.5] {$x^k(1)$}   (MLM_Tx_0);
    \draw[arrow,thin] (MLM_Mid_Block) |- node[left, pos=.5] {$x^k(B^k)$} (MLM_Tx_last);

    \node[coord, right = 17mm of MLM_Tx_0] (Interleaver_in_0) {};
    \node[coord, right = 17mm of MLM_Tx_last] (Interleaver_in_last) {};
    \node[block, right = 17mm of MLM_Tx_dots, minimum height=42mm] (Interleaver) {$\Pi$};
    \draw[arrow,thin] (MLM_Tx_0) -- node[above, pos=.485] {$m^k_i(1)$}   (Interleaver_in_0);
    \draw[arrow,thin] (MLM_Tx_last) -- node[above, pos=.485] {$m^k_i(B^k)$} (Interleaver_in_last);

     \node[coord, right = 18mm of Interleaver_in_0] (Interleaver_out_0) {};
    \node[coord, right = 18mm of Interleaver_in_last] (Interleaver_out_last) {};

    \node[block, right = 20mm of Interleaver_out_0] (Gauss_0) {$H_i$};
    \node[block, right = 20mm of Interleaver, fill=none, draw=none] (Gauss_mid) {$\bullets$};
    \node[block, right = 20mm of Interleaver_out_last] (Gauss_last) {$H_i$};
    \draw[arrow,thin] (Interleaver_out_0) -- node[above, pos=.485] {$\grave{m}^B_{i;(1,1)}$}   (Gauss_0);
    \draw[arrow,thin] (Interleaver_out_last) -- node[above, pos=.485] {$\grave{m}^B_{i;(k,B^{k-1})}$} (Gauss_last);
    
    \node[coord, right = 12mm of Gauss_mid, fill=none, draw=none] (Output_mid) {$\bullets$};
    \node[coord, right = 12mm of Gauss_0, fill=none, draw=none] (Output_0) {};
    \node[coord, right = 12mm of Gauss_last, fill=none, draw=none] (Output_last) {};
    \draw[arrow,thin] (Gauss_0) -- node[above, pos=.9] {$\tilde{m}^B_{i;(1,1)}$} (Output_0);
    \draw[arrow,thin] (Gauss_last) -- node[above, pos=.9] {$\tilde{m}^B_{i;(k,B^{k-1})}$} (Output_last);

    \node[block, below = 5mm of Output_0, fill=none, draw=none] (SI_0) {};
    \node[block, below = 6mm of Output_last, fill=none, draw=none] (SI_last) {};
    
    \begin{pgfonlayer}{background}
        \node[draw, rounded corners, dashed, fit = (input_for_frame) (Gauss_last) (MLM_Tx_0)] (Tx) {};
    \end{pgfonlayer}
    \node[above = 0 of Tx] {}; 

\end{tikzpicture}
    }
    \caption{Transmitter}
\end{subfigure}
\begin{subfigure}[t]{0.5\columnwidth}
	\centering
    \resizebox{\textwidth}{!}{\pgfmathsetseed{4}

\begin{tikzpicture}[auto, arrow/.style={very thick, ->, >=stealth'},node distance=5mm,>=latex']
    \node [coord] (input) {};
    \node [coord, right = 28mm of input] (input_for_frame) {};

    
    \node[coord, right = 6mm of input] (Base_DeGauss_node) {};
    
    \node[block, above right = 10mm and 14mm of Base_DeGauss_node, align = center] (DeGauss0)
    {$H^{\dagger}_i$};
    \node[coord, left = 21mm of DeGauss0] (DeGauss0_input) {};
    \node[block, below = 5mm of DeGauss0, fill=none, draw=none, align = center] (DeGauss_dots) {$\bullets$};
    \node[block, below = 5mm of DeGauss_dots, align = center] (DeMLM_Rx_last) {$H^{\dagger}_i$};
     \node[coord, left = 21mm of DeMLM_Rx_last] (DeMLM_Rx_last_input) {};
    \node[coord, right = 5mm of Base_DeGauss_node, fill=none, draw=none, align = center] (DeGauss_Mid_Block) {};
    \draw[arrow,thin] (DeGauss0_input) -- node[above, pos=.5] {\small $\hat{\tilde{m}}^B_{i;1}(1)$}   (DeGauss0);
    \draw[arrow,thin] (DeMLM_Rx_last_input) -- node[above, pos=.5] {\small $\hat{\tilde{m}}^B_{i;k}\left(B^{k - 1}\right)$} (DeMLM_Rx_last);

    \node[coord, right = 19mm of DeGauss0] (Interleaver_in_0) {};
    \node[coord, right = 19mm of DeMLM_Rx_last] (Interleaver_in_last) {};
    \node[block, right = 19mm of DeGauss_dots, minimum height=42mm, minimum width=17mm] (Interleaver) {$\Pi^{-1}$};
    \draw[arrow,thin] (DeGauss0) -- node[above, pos=.485] {\small $\hat{\grave{m}}^B_{i;1}(1)$}   (Interleaver_in_0);
    \draw[arrow,thin] (DeMLM_Rx_last) -- node[above, pos=.485] {\small $\hat{\grave{m}}^B_{i;k}\left(B^{k-1}\right)$} (Interleaver_in_last);

     \node[coord, right = 17mm of Interleaver_in_0] (Interleaver_out_0) {};
    \node[coord, right = 17mm of Interleaver_in_last] (Interleaver_out_last) {};

    \node[block, right = 20mm of Interleaver_out_0] (MLM_Rx_0) {\begin{tabular}{c} MLM \\ Rx$_i$ \end{tabular}};
    \node[block, right = 20mm of Interleaver, fill=none, draw=none] (Gauss_mid) {$\bullets$};
    \node[block, right = 20mm of Interleaver_out_last] (MLM_Rx_last) {\begin{tabular}{c} MLM \\ Rx$_i$ \end{tabular}};
    \draw[arrow,thin] (Interleaver_out_0) -- node[above, pos=.485] {$\hat{m}^k_i(1)$}   (MLM_Rx_0);
    \draw[arrow,thin] (Interleaver_out_last) -- node[above, pos=.485] {$\hat{m}^k_i\left(B^{k}\right)$} (MLM_Rx_last);
    
    \node[coord, right = 12mm of Gauss_mid, fill=none, draw=none] (Output_mid) {$\bullets$};
    \node[coord, right = 12mm of MLM_Rx_0, fill=none, draw=none] (Output_0) {};
    \node[coord, right = 12mm of MLM_Rx_last, fill=none, draw=none] (Output_last) {};
    \draw[arrow,thin] (MLM_Rx_0) -- node[above, pos=.9] {$\hat{x}^{k}_{i}(1)$} (Output_0);
    \draw[arrow,thin] (MLM_Rx_last) -- node[above, pos=.9] {$\hat{x}^{k}_{i}\left(B^{k}\right)$} (Output_last);
    \node[block, below = 5mm of Output_0, fill=none, draw=none] (SI_0) {$\hat{x}^{k}_{i-1}(1)$};
    \node[block, below = 6mm of Output_last, fill=none, draw=none] (SI_last) {$\hat{x}^{k}_{i-1}\left(B^{k}\right)$};
    \draw[arrow,thin] (SI_0) -| (MLM_Rx_0);
    \draw[arrow,thin] (SI_last) -| (MLM_Rx_last); 
    
    \node [coord, below = 0.5mm of MLM_Rx_last] (Output_frame_lower_border) {};

    \begin{pgfonlayer}{background}
        \node[draw, rounded corners, dashed, fit = (input_for_frame) (MLM_Rx_last) (MLM_Rx_0) (Output_frame_lower_border) (DeGauss0)] (Tx) {};
    \end{pgfonlayer}
    \node[above = 0 of Tx] {}; 

\end{tikzpicture}
    }
    \caption{Receiver} 
\end{subfigure}
\caption{Block diagram for the i'th MLM layer transmitter and receiver of \schemeref{schm:UniversalAnalogMLM_Full}. We denote the interleaving and deinterleaving operations by $\Pi$ and $\Pi^{-1}$, respectively.}
\label{fig:SubBlocks_PiTxRx}
\end{figure}
}
{
\begin{figure}
\begin{subfigure}[b]{\columnwidth}
	\centering
    \resizebox{\textwidth}{!}{\pgfmathsetseed{4}

\begin{tikzpicture}[auto, arrow/.style={very thick, ->, >=stealth'},node distance=5mm,>=latex']
    \node [coord] (input) {};
    \node [coord, right = 25mm of input] (input_for_frame) {};

    
    \node[coord, right = 6mm of input] (Base_MLM_node) {};
    \node[block, above right = 10mm and 8mm of Base_MLM_node, align = center] (MLM_Tx_0) {MLM \\ Tx$_i$};
    \node[block, below = 5mm of MLM_Tx_0, fill=none, draw=none, align = center] (MLM_Tx_dots) {$\bullets$};
    \node[block, below = 5mm of MLM_Tx_dots, align = center] (MLM_Tx_last) {MLM \\ Tx$_i$};
    \node[coord, right = 5mm of Base_MLM_node, fill=none, draw=none, align = center] (MLM_Mid_Block) {};
    \draw[arrow] (input) -- node[above, pos=.03] {$\mX$} (MLM_Mid_Block);
    \draw[arrow,thin] (MLM_Mid_Block) |- node[left, pos=.5] {$x^k(1)$}   (MLM_Tx_0);
    \draw[arrow,thin] (MLM_Mid_Block) |- node[left, pos=.5] {$x^k(B^k)$} (MLM_Tx_last);

    \node[coord, right = 17mm of MLM_Tx_0] (Interleaver_in_0) {};
    \node[coord, right = 17mm of MLM_Tx_last] (Interleaver_in_last) {};
    \node[block, right = 17mm of MLM_Tx_dots, minimum height=42mm] (Interleaver) {$\Pi$};
    \draw[arrow,thin] (MLM_Tx_0) -- node[above, pos=.485] {$m^k_i(1)$}   (Interleaver_in_0);
    \draw[arrow,thin] (MLM_Tx_last) -- node[above, pos=.485] {$m^k_i(B^k)$} (Interleaver_in_last);

     \node[coord, right = 18mm of Interleaver_in_0] (Interleaver_out_0) {};
    \node[coord, right = 18mm of Interleaver_in_last] (Interleaver_out_last) {};

    \node[block, right = 20mm of Interleaver_out_0] (Gauss_0) {$H_i$};
    \node[block, right = 20mm of Interleaver, fill=none, draw=none] (Gauss_mid) {$\bullets$};
    \node[block, right = 20mm of Interleaver_out_last] (Gauss_last) {$H_i$};
    \draw[arrow,thin] (Interleaver_out_0) -- node[above, pos=.485] {$\grave{m}^B_{i;(1,1)}$}   (Gauss_0);
    \draw[arrow,thin] (Interleaver_out_last) -- node[above, pos=.485] {$\grave{m}^B_{i;(k,B^{k-1})}$} (Gauss_last);
    
    \node[coord, right = 12mm of Gauss_mid, fill=none, draw=none] (Output_mid) {$\bullets$};
    \node[coord, right = 12mm of Gauss_0, fill=none, draw=none] (Output_0) {};
    \node[coord, right = 12mm of Gauss_last, fill=none, draw=none] (Output_last) {};
    \draw[arrow,thin] (Gauss_0) -- node[above, pos=.9] {$\tilde{m}^B_{i;(1,1)}$} (Output_0);
    \draw[arrow,thin] (Gauss_last) -- node[above, pos=.9] {$\tilde{m}^B_{i;(k,B^{k-1})}$} (Output_last);

    \node[block, below = 5mm of Output_0, fill=none, draw=none] (SI_0) {};
    \node[block, below = 6mm of Output_last, fill=none, draw=none] (SI_last) {};
    
    \begin{pgfonlayer}{background}
        \node[draw, rounded corners, dashed, fit = (input_for_frame) (Gauss_last) (MLM_Tx_0)] (Tx) {};
    \end{pgfonlayer}
    \node[above = 0 of Tx] {}; 

\end{tikzpicture}
    }
    \caption{Interleaved MLM Tx$_i$ ($\Pi$MLM$_i$)}
\end{subfigure}
\\ \ \\
\begin{subfigure}[b]{\columnwidth}
	\centering
    \resizebox{\textwidth}{!}{\pgfmathsetseed{4}

\begin{tikzpicture}[auto, arrow/.style={very thick, ->, >=stealth'},node distance=5mm,>=latex']
    \node [coord] (input) {};
    \node [coord, right = 28mm of input] (input_for_frame) {};

    
    \node[coord, right = 6mm of input] (Base_DeGauss_node) {};
    
    \node[block, above right = 10mm and 14mm of Base_DeGauss_node, align = center] (DeGauss0)
    {$H^{\dagger}_i$};
    \node[coord, left = 21mm of DeGauss0] (DeGauss0_input) {};
    \node[block, below = 5mm of DeGauss0, fill=none, draw=none, align = center] (DeGauss_dots) {$\bullets$};
    \node[block, below = 5mm of DeGauss_dots, align = center] (DeMLM_Rx_last) {$H^{\dagger}_i$};
     \node[coord, left = 21mm of DeMLM_Rx_last] (DeMLM_Rx_last_input) {};
    \node[coord, right = 5mm of Base_DeGauss_node, fill=none, draw=none, align = center] (DeGauss_Mid_Block) {};
    \draw[arrow,thin] (DeGauss0_input) -- node[above, pos=.5] {\small $\hat{\tilde{m}}^B_{i;1}(1)$}   (DeGauss0);
    \draw[arrow,thin] (DeMLM_Rx_last_input) -- node[above, pos=.5] {\small $\hat{\tilde{m}}^B_{i;k}\left(B^{k - 1}\right)$} (DeMLM_Rx_last);

    \node[coord, right = 19mm of DeGauss0] (Interleaver_in_0) {};
    \node[coord, right = 19mm of DeMLM_Rx_last] (Interleaver_in_last) {};
    \node[block, right = 19mm of DeGauss_dots, minimum height=42mm, minimum width=17mm] (Interleaver) {$\Pi^{-1}$};
    \draw[arrow,thin] (DeGauss0) -- node[above, pos=.485] {\small $\hat{\grave{m}}^B_{i;1}(1)$}   (Interleaver_in_0);
    \draw[arrow,thin] (DeMLM_Rx_last) -- node[above, pos=.485] {\small $\hat{\grave{m}}^B_{i;k}\left(B^{k-1}\right)$} (Interleaver_in_last);

     \node[coord, right = 17mm of Interleaver_in_0] (Interleaver_out_0) {};
    \node[coord, right = 17mm of Interleaver_in_last] (Interleaver_out_last) {};

    \node[block, right = 20mm of Interleaver_out_0] (MLM_Rx_0) {\begin{tabular}{c} MLM \\ Rx$_i$ \end{tabular}};
    \node[block, right = 20mm of Interleaver, fill=none, draw=none] (Gauss_mid) {$\bullets$};
    \node[block, right = 20mm of Interleaver_out_last] (MLM_Rx_last) {\begin{tabular}{c} MLM \\ Rx$_i$ \end{tabular}};
    \draw[arrow,thin] (Interleaver_out_0) -- node[above, pos=.485] {$\hat{m}^k_i(1)$}   (MLM_Rx_0);
    \draw[arrow,thin] (Interleaver_out_last) -- node[above, pos=.485] {$\hat{m}^k_i\left(B^{k}\right)$} (MLM_Rx_last);
    
    \node[coord, right = 12mm of Gauss_mid, fill=none, draw=none] (Output_mid) {$\bullets$};
    \node[coord, right = 12mm of MLM_Rx_0, fill=none, draw=none] (Output_0) {};
    \node[coord, right = 12mm of MLM_Rx_last, fill=none, draw=none] (Output_last) {};
    \draw[arrow,thin] (MLM_Rx_0) -- node[above, pos=.9] {$\hat{x}^{k}_{i}(1)$} (Output_0);
    \draw[arrow,thin] (MLM_Rx_last) -- node[above, pos=.9] {$\hat{x}^{k}_{i}\left(B^{k}\right)$} (Output_last);
    \node[block, below = 5mm of Output_0, fill=none, draw=none] (SI_0) {$\hat{x}^{k}_{i-1}(1)$};
    \node[block, below = 6mm of Output_last, fill=none, draw=none] (SI_last) {$\hat{x}^{k}_{i-1}\left(B^{k}\right)$};
    \draw[arrow,thin] (SI_0) -| (MLM_Rx_0);
    \draw[arrow,thin] (SI_last) -| (MLM_Rx_last); 
    
    \node [coord, below = 0.5mm of MLM_Rx_last] (Output_frame_lower_border) {};

    \begin{pgfonlayer}{background}
        \node[draw, rounded corners, dashed, fit = (input_for_frame) (MLM_Rx_last) (MLM_Rx_0) (Output_frame_lower_border) (DeGauss0)] (Tx) {};
    \end{pgfonlayer}
    \node[above = 0 of Tx] {}; 

\end{tikzpicture}
    }
    \caption{Deinterleaved MLM Rx$_i$ ($\Pi^{-1}$MLM$_i$)} 
\end{subfigure}
\caption{Block diagram for the i'th MLM layer transmitter and receiver of \schemeref{schm:UniversalAnalogMLM_Full}. We denote the interleaving and deinterleaving operations by $\Pi$ and $\Pi^{-1}$, respectively.}
\label{fig:SubBlocks_PiTxRx}
\end{figure}
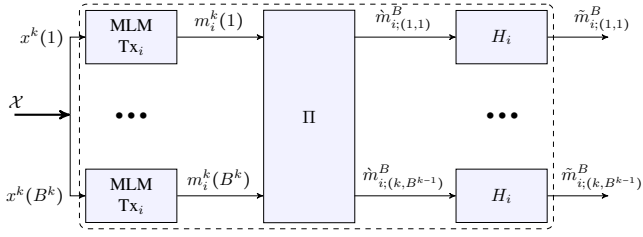
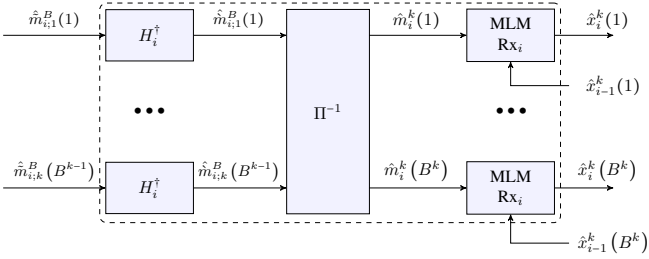
}
\begin{scheme}[Full MLM-based]\ \\
 \label{schm:UniversalAnalogMLM_Full}
    \textit{\textbf{$M$-Layer Transmitter:}}
    
    \textit{First layer ($i=1$):}
    \begin{itemize}
    \item 
        For $B \geq k$, $B \in \nats$, accumulates $B^k$ source (column) vectors $x^k(1), x^k(2), \ldots, x^k(B^k)$. Denote by $\mX$ the matrix whose columns are the source vectors:
        \begin{align}
            \mX \triangleq \begin{bmatrix} x^k(1) & x^k(2) & \ldots & x^k(B^k) \end{bmatrix}.
        \end{align}
    \item 
        For each $b \in \{1, 2, \ldots, B^k\}$, transmits each of the entries of the vector $x^k(b)$ over the channel \eqref{eq:ChannelEq} linearly \eqref{eq:Background_KnownENR_DirectLinear}:
        \begin{align}
        \label{eq:MLM:UnknownENR:Tx_Layer1_Full}
            s_{1;\ell,b}(t) &\triangleq s\left(t+\left(\ell-1\right) T + \left(b-1\right) kT\right)
         \col{}{\\ &}=
            \sqrt{\frac{E_1}{T}} \frac{x_\ell(b)}{\sigma_x} \varphi(t), 
            \qquad t \in [0, T), 
        \end{align}
        for $\ell = \{1, 2, \ldots, k\}$, where $\varphi$ 
        is a 
        continuous 
        unit-norm (i.e., unit-energy) 
        waveform that is zero outside the interval $[0,T]$, say $\phi$ of \eqref{eq:AnalogPPM:PulseShaping_RectPulse}, 
        $E_1 \in [0, E]$ is the allocated energy for layer $1$, and $E$ is the total available energy of the scheme.
    \end{itemize}

    \textit{Other layers:} For each $i \in \{2, \ldots, M\}$:
        \begin{itemize}
        \item 
            For each $b \in \{1, 2, \ldots, B^k\}$, calculates the $k$-dimensional tuple 
            \begin{align}
            \label{eq:RepeatedQuant_Tx_Full}
                m^k_i(b) &= [\eta_i(b) x^k(b) + d^k_i(b)]_\Lambda \,, 
            \end{align}
            where 
            $m^k_i(b) = \left(m_{i;1}(b), m_{i;2}(b), \ldots, m_{i;k}(b) \right)^\dagger$, 
            and $m_{i;\ell}(b)$ denotes the 
            $\ell^\text{th}$ entry of $m^k_i(b)$
            for $\ell \in \{1, \ldots k\}$; 
            $\eta_i(b)$, $d^k_i(b)$ and $\Lambda$ take the roles of $\eta,d^k$ and $\Lambda$ of \schemeref{schm:MLM_Reznic}, and are tailored for each layer $i$; $\Lambda$ is chosen to have unit second moment. 
        \item 
            For each $\ell \in \{1, \ldots, k\}$, 
            interleaves the entries $m_{i;\ell}(1), \ldots, m_{i;\ell}(B^k)$, 
            stacks them into vectors of size $B$, and applies to each of them a $B$-dimensional orthogonal matrix $\mH_i$, as follows.
            
            \begin{subequations}
            \label{eq:interleave}
            \noeqref{eq:interleave:Gaussianize:compact,eq:interleave:Gaussianize}
            \begin{align}
                \col{}{&}\tm^{B}_{i;(\ell,j)} 
        \col{}{\\* &\!} = \mH_i 
                \begin{pmatrix}
                    m_{i;\ell} \left(\left\lfloor \frac{j-1}{B^\ell} \right\rfloor \cdot B^{\ell + 1} +  [j-1]_{B^\ell} + 1\right)
                 \\ m_{i;\ell} \left(\left\lfloor \frac{j-1}{B^\ell} \right\rfloor \cdot B^{\ell + 1} +  [j-1]_{B^\ell} + B^\ell + 1\right)
                 \\ m_{i;\ell} \left(\left\lfloor \frac{j-1}{B^\ell} \right\rfloor \cdot B^{\ell + 1} +  [j-1]_{B^\ell} + 2B^\ell + 1 \right)
                 \\ \vdots
                 \\ m_{i;\ell} \left(\left\lfloor \frac{j-1}{B^\ell} \right\rfloor \cdot B^{\ell + 1} +\ [j-1]_{B^\ell} + B^{\ell}\left(B -1\right) + 1\right)
                \end{pmatrix}
            \label{eq:interleave:Gaussianize}
         \col{}{\\& \!}
            \triangleq
                \mH_i \grave{m}^{B}_{i;(\ell,j)}
            \col{}{\label{eq:interleave:Gaussianize:compact}}
            \end{align}
            \end{subequations}
            for $j \in \{1, 2, \ldots, B^{k - 1}\}$, 
            where 
            $\grave{m}^{B}_{i;(\ell,j)}$ is the vector after interleaving;
            $\tm^{B}_{i;(\ell,j)}$ is the vector after interleaving and matrix multiplication and its $\xi^\text{th}$ entry is $\tm_{i;\xi,(\ell,j)}$ for $\xi \in \{1,\ldots, B\}$; the length of the vectors 
            $\grave{m}^{B}_{i;(\ell,j)}$ and $\tm^{B}_{i;(\ell,j)}$ is $B$. Note that the interleaving operation creates doubly-indexed vectors, where a set of $B^k$ vectors of length $k$ is transformed into $k \times B^{k-1}$ vectors of length $B$, which are indexed by $\ell \in \{ 1,\ldots, k \}$ and $j \in \left\{ 1,\ldots, B^{k-1} \right\}$.
           
         \item 
            For each $\ell$, $j$, and $\xi$, 
            views $\tm_{i;\xi,(\ell,j)}$ 
            as a scalar source sample, 
            and generates a corresponding channel input $\left\{ s_{i;\xi,(\ell,j)}(t) \middle| t \in [0,T) \right\}$ where 
            \begin{align}
                &s_{i;\xi,(\ell,j)}(t) 
             \col{}{\\ & }\triangleq\! s \big(t + \big((\ell-1) + (\xi-1) k + (j-1) Bk + (i-1) k B^k \big) T \big) 
            \nonumber
            \end{align}
            using a scalar JSCC scheme with a predefined energy $E_i \geq 0$ that is designed for a predetermined $\ENR_i$, or equivalently, $N_{i} = E_i/\ENR_i$, such that $\sum_{i=1}^{M} E_i = E$ and $N_2 > N_3 > \cdots > N_{M} > 0$. 
            
         \end{itemize}

\vspace{.5\baselineskip}
    \textit{\textbf{Receiver:}} Receives the channel output signal $r$ \eqref{eq:ChannelEq} and recovers the different layers as follows.

    \textit{First layer ($i=1$):}
        For each $\ell \in \{1, \ldots, k\}$, $b \in \{1, \ldots, B^k\}$:
        \begin{itemize}
        \item 
            Recovers the MMSE estimate $\hx_{1;\ell}(b)$ of $x_\ell(b)$ given
            $\left\{ r_{1;\ell,b}(t) | t \in [0, T) \right\}$, 
            where
            \begin{align}
                r_{1;\ell,b}(t) \triangleq r \left( t + (\ell-1) T + (b-1) k T \right) .
            \end{align}
            Denote the matrix whose columns comprise these estimates by 
            $\hmX_1 \triangleq  
            \begin{bmatrix} \hx^{k}_{1}(1) & \ldots & \hx^{k}_{1}(B^k) \end{bmatrix}$. 
        \item 
            If the true noise level $N$ satisfies $N > N_2$, sets the final estimate 
            $\hmX$ of $\mX$ to $\hmX_1$ 
            and stops. Otherwise, determines the maximal layer index $\jmath$ for which $N \leq N_\jmath$ and continues to process the other layers. 
        \end{itemize}
        
    \textit{Other layers:}
        For each $i \in \{ 2, \ldots, \jmath \}$ in ascending order:
        \begin{itemize}
        \item 
            For each $\ell \in \{1, \ldots, k\}, \ j \in \{1, \ldots, B^{k-1}\}$ and $\xi \in \{1, \ldots, B\}$, uses the receiver of the scalar JSCC scheme to generate an estimate 
            $\htm^{B}_{i;(\ell,j)}$ of $\tm^{B}_{i;(\ell,j)}$
            from $\left\{ r_{i;\xi,(\ell,j)}(t) \middle| t \in [0, T) \right\}$, where
            \begin{align}
               &r_{i;\xi,(\ell,j)}(t) \col{}{\\ &}\triangleq 
               r\left(t + \left((\ell - 1) + (\xi - 1) k + (j - 1) Bk + (i - 1) k B^k\right)T\right) .
           \nonumber
            \end{align}
            
        \item
            For each $\ell \in \{1, \ldots, k\}$, 
            stacks the entries of $\htm^{B}_{i;(\ell,1)}, \ldots, \htm^{B}_{i;(\ell,B^{k-1})}$ into vectors of length $B$, $\htm^{B}_{i;(\ell,j)}$, applies the orthogonal matrix $\mH_i^{-1} = \mH_i^\dagger$ to each vector $\htm^{B}_{i;(\ell,j)}$, and deinterleaves the outcomes, to attain $\hat{\grave{m}}^{B}_{i;(\ell,j)}$, as follows.
            
           \begin{subequations}
            \label{eq:deinterleave}
            \noeqref{eq:deinterleave:deinterleave,eq:deinterleave:Gaussianize}
            \begin{align}
                &\hat{\grave{m}}^{B}_{i;(\ell,j)} 
             \col{}{\\* &}=
                \begin{pmatrix}
                    \hm_{i;\ell} \left(\left\lfloor \frac{j-1}{B^\ell} \right\rfloor \cdot B^{\ell + 1} +  [j-1]_{B^\ell} \right)
                 \\ \hm_{i;\ell} \left(\left\lfloor \frac{j-1}{B^\ell} \right\rfloor \cdot B^{\ell + 1} +  [j-1]_{B^\ell} + B^\ell \right)
                 \\ \hm_{i;\ell} \left(\left\lfloor \frac{j-1}{B^\ell} \right\rfloor \cdot B^{\ell + 1} +  [j-1]_{B^\ell} + 2B^\ell \right)
                 \\ \vdots
                 \\ \hm_{i;\ell} \left(\left\lfloor \frac{j-1}{B^\ell} \right\rfloor \cdot B^{\ell + 1} +  [j-1]_{B^\ell} + B^{\ell}\left(B -1\right) \right)
                \end{pmatrix}
            \col{}{\qquad
            \label{eq:deinterleave:deinterleave}
             \\ &}= \mH_i^\dagger \htm^{B}_{i;(\ell,j)}
            \label{eq:deinterleave:Gaussianize}
            \end{align}
           \end{subequations}
                for $j \in \left\{1, 2, \ldots, B^{k-1}\right\}$.
        \item
            For each $b \in \left\{1, \ldots, B^k\right\}$,
            using the effective channel output $\hm_i^k(b)$ (that takes the role of $y^k$ in \schemeref{schm:MLM_Reznic}) with SI $\hx_{i-1}^k(b)$, generates the signal 
            \begin{align}
                \label{eq:MLM:UnknownENR:Layers_Full}
                \ty^k_i(b) &= [\alpha^{(i)}_c \hm^k_i(b) - \eta_i \hx^k_{i-1}(b) - d^k_i(b)]_\Lambda,
            \end{align}
            as in \eqref{eq:MLM_Rx1} of \schemeref{schm:MLM_Reznic}, 
            where $\alpha^{(i)}_c$ is a channel scale factor.
        \item 
            For each $b \in \left\{1, \ldots, B^k\right\}$, 
            constructs an estimate $\hx^k(b)$ of $x^k(b)$:
            \begin{align}
                \hx_i^k(b) &= \frac{\alpha^{(i)}_s}{\eta_i}\ty^k_i(b) + \hx^k_{i-1}(b),
            \end{align}
            as in \eqref{eq:MLM_Rx2} of \schemeref{schm:MLM_Reznic}, 
            where $\alpha^{(i)}_s$ is a source scale factor.
            Denote the matrix whose columns comprise these estimates by 
            $\hmX_i \triangleq  
            \begin{bmatrix} \hx^{k}_{i}(1) & \ldots & \hx^{k}_{i}(B^k) \end{bmatrix}$. 
            The final estimate is $\hmX = \hmX_\jmath$.
     \end{itemize}
 \end{scheme} 


\section{Proof of \thmref{thm:linear_MLM}}
\label{app:proof:thm:linear_MLM}

To prove \thmref{thm:linear_MLM}, we will make use of the following lemma about the validity of the MLM results from \secref{s: MLM} for the multi-layer MLM scenario, where the mid-stage noise vectors are linear combinations of dithers and Gaussian noises. The proof of this lemma is given in \appref{app:error:semi-norm-ergodic}.  
\begin{lem}
\label{lem:error:semi-norm-ergodic}
    Let $q^k$ be a sequence in $k$ of vectors, such that the vector $q^k$ equals with probability $1 - P_k$ to a linear combination of a Gaussian vector and dithers all of which are mutually independent, where $\lim_{k \to \infty} P_k = 0$. 
    Then, the sequence in $k$ of error signals $x^k - \hx^k$ is SNE
    for a sequence of lattices that is good for both channel coding and MSE quantization; moreover, for each $k$, the error signal equals with probability $1 - Q_k$ to a linear combination of a Gaussian vector and dithers all of which are mutually independent, where $\lim_{k \to \infty} Q_k = 0$. 
\end{lem}

We now prove \thmref{thm:linear_MLM}.
    We will construct a scheme with a large enough (yet finite) $M$ that achieves \eqref{eq:Profile_Polynom} with the predefined $L$ and $\Edsgn$ for all $N > \Nmin$ for a given $\Nmin > 0$.
    For any $\Nmin > 0$, however small, we will choose $M \in \nats$ large enough and $\{N_i | i = 1, \ldots, M\}$ such that $\Nmin \in (N_M, N_{M-1}]$. 
    
    Consider the first layer ($i=1$). The distortion $D_1$ of $\hx_1^k$ for a noise level $N$ is bounded from above by
    
    \begin{subequations}
    \label{eq:proof:unknown-ENR:D0}
    \begin{align}
        D_1(N) 
        &= \frac{\sigma^2_x}{1 + \frac{2 E_1}{N}}
    \label{eq:proof:unknown-ENR:D0:SLB}
     \\ &\leq \sigma_x^2 \cdot \mF(N)
    \label{eq:proof:unknown-ENR:D0:profile}
     \\ &= \frac{\sigma_x^2}{1 + \left(\frac{\Edsgn}{N}\right)^L}
    \label{eq:proof:unknown-ENR:D0:explicit}
    \end{align}
    \end{subequations}
    where \eqref{eq:proof:unknown-ENR:D0:SLB} follows from \eqref{eq:Background_KnownENR_DirectLinearMSE},
    and \eqref{eq:proof:unknown-ENR:D0:profile} and \eqref{eq:proof:unknown-ENR:D0:explicit} follow from the distortion profile requirement \eqref{eq:profile:tot} for $N > N_2$.
    
    To guarantee the requirement \eqref{eq:proof:unknown-ENR:D0:profile} for all $N > N_2$, 
    it suffices to guarantee it for the extreme value $N=N_2$, 
    which holds, in turn, for 
    \begin{align}
    \label{eq:FirstLayerEnergy}
        E_1 &=  \left( \frac{\Edsgn}{N_2} \right)^L \frac{N_2}{2}.
    \end{align}    
    
    For $i \in \{2, \ldots, \jmath\}$, 
    the distortion $D_i$ of $\hx_i^k$ for a noise level $N$ is bounded from above by 
    \begin{subequations} 
    \label{eq:proof:unknown-ENR:Di}
    \begin{align}
        D_i(N) 
        &\leq \frac{D_{i - 1}(N_{i})}{1 + \frac{2E_i}{N}}\cdot\frac{1 + \frac{2E_i}{N_{i}}}{\frac{2E_i}{N_{i}}} + \eps_i
    \label{eq:proof:unknown-ENR:Di:recursion}
     \\ &\leq \frac{\sigma_x^2 \cdot \mF(N_{i})}{1 + \frac{2E_i}{N}}\cdot\frac{1 + \frac{2E_i}{N_{i}}}{\frac{2E_i}{N_{i}}} + \eps_i
    \label{eq:proof:unknown-ENR:Di:profile-prev}
     \\ &= \frac{\sigma_x^2}{1 + \left(\frac{\Edsgn}{N_{i}}\right)^L} \cdot \frac{1}{1 + \frac{2E_i}{N}}\cdot\frac{1 + \frac{2E_i}{N_{i}}}{\frac{2E_i}{N_{i}}} + \eps_i
    \label{eq:proof:unknown-ENR:Di:profile-prev-explicit}
     \\ &\leq \sigma_x^2 \cdot \mF(N)
    \label{eq:proof:unknown-ENR:Di:profile-new}
     \\ &= \frac{\sigma_x^2}{1 + \left(\frac{\Edsgn}{N}\right)^L} ,
    \label{eq:proof:unknown-ENR:Di:profile-new-explicit}
    \end{align}
    \end{subequations} 
    where 
    \eqref{eq:proof:unknown-ENR:Di:recursion} follows from \corref{cor:SNR_Universal_MLM}
    by treating $\hx_{i-1}^k$ as SI and the error $x - \hx_{i-1}^k$ taking the role of the ``unknown part'' at the receiver with power $D_{i-1}$, with $\eps_i$ going to zero with $k$, and by invoking \lemref{lem:error:semi-norm-ergodic} recursively, which guarantees that the sequence in $k$ of the error vectors $x - \hx_{i-1}^k$ is SNE;
    \eqref{eq:proof:unknown-ENR:Di:profile-prev} holds by the distortion profile requirement \eqref{eq:profile:tot};\footnote{The requirement $D_{i-1}(N) \leq \sigma_x^2 \cdot \mF(N)$ is satisfied for $N = N_i - \eps$ for any $\eps > 0$, however small, and therefore, holds also for $N = N_i$, by continuity. Alternatively, one may view it as a requirement of the scheme given $i-1$ layers, for all $i \in \{2, 3, \ldots, \jmath\}$.} 
    \eqref{eq:proof:unknown-ENR:Di:profile-prev-explicit} follows from \eqref{eq:Profile_Polynom} with $\Edsgn$ and $L$;
    \eqref{eq:proof:unknown-ENR:Di:profile-new} follows from the distortion profile requirement \eqref{eq:profile:tot} for $N \in (N_{i+1}, N_{i}]$;
    and \eqref{eq:proof:unknown-ENR:Di:profile-new-explicit} follows from \eqref{eq:Profile_Polynom} with $\Edsgn$ and $L$.
    
    To guarantee the requirement \eqref{eq:proof:unknown-ENR:Di:profile-new} for all $N \in (N_{i+1}, N_{i}]$
    we need only to satisfy it for the extreme value $N=N_{i+1}$, 
    which holds, in turn, for 
    \begin{subequations}
    \label{eq:MidEnergy_Linear}
    \noeqref{eq:MidEnergy_Linear:basic} 
    \begin{align}
        1 + \frac{2E_i}{N_{i+1}} &\geq \frac{1 + \frac{2E_i}{N_{i}}}{\frac{2 E_i}{N_{i}}} \cdot \frac{1 + \left( \frac{\Edsgn}{N_{i+1}} \right)^L}{1 + \left( \frac{\Edsgn}{N_{i}} \right)^L} + \teps_i 
    \label{eq:MidEnergy_Linear:basic}
     \\ &\geq \left(1 + \frac{N_{i}}{2E_i}\right)\left(\frac{N_{i}}{N_{i+1}}\right)^{L} + \teps_i ,
    \label{eq:MidEnergy_Linear:drop1s}
    \end{align}
    \end{subequations}
    where \eqref{eq:MidEnergy_Linear:drop1s} holds since $N_{i+1} < N_{i}$. and $\teps_i$ decays to zero with $k$;
    the set of inequalities \eqref{eq:MidEnergy_Linear} holds for 
    \begin{align}
    \label{eq:Energy_Linear_Final}
        E_i &= \frac{N_{i+1}}{4}\left(\left(\frac{N_{i}}{N_{i+1}}\right)^L\! - 1\right)\left(1 +\! \sqrt{1 + \frac{4\left(\frac{N_{i}}{N_{i+1}}\right)^{L + 1}}{\left(1 - \left(\frac{N_{i}}{N_{i+1}}\right)^L\right)^2}}\right) + \veps_i, 
    \end{align}
    where again $\veps_i$ decay to zero with $k$.
    
    We are now ready to bound the total energy $E$.
    \begin{subequations}
    \label{eq:proof:unknown-ENR:Etot}
    \noeqref{eq:proof:unknown-ENR:Etot:def}
    \begin{align}
        &\frac{E}{\Edsgn} = \frac{1}{\Edsgn} \sum_{i = 1}^{\jmath} E_{i}
    \label{eq:proof:unknown-ENR:Etot:def}
     \\ & \leq \sum_{i = 2}^{\infty}\frac{N_{i+1}}{4 \Edsgn}\left(\left(\frac{N_{i}}{N_{i+1}}\right)^L - 1\right)\left(1 + \sqrt{1 + \frac{4\left(\frac{N_{i}}{N_{i+1}}\right)^{L + 1}}{\left(1 - \left(\frac{N_{i}}{N_{i+1}}\right)^L\right)^2}}\right)
    \col{}{\nonumber 
    \\* &\qquad\qquad} + \frac{1}{2} \left( \frac{\Edsgn}{N_2} \right)^{L-1} + \sum_{i=2}^{\jmath} \frac{\veps_i}{\Edsgn}
    \label{eq:proof:unknown-ENR:Etot:substitute}
     \\ &= \frac{\Delta}{4\Edsgn} \left(\e^{\alpha L} - 1\right)\left(1 + \sqrt{1 + \frac{4\e^{\alpha\left(L + 1\right)}}{\left(1 - \e^{\alpha L}\right)^2}}\right)\sum_{i = 2}^{\infty}\e^{-\alpha i}
    \col{}{\\* &\qquad\qquad} + \half \left(\frac{\Edsgn}{\Delta\e^{-\alpha}}\right)^{L - 1} + \sum_{i=2}^{\jmath} \frac{\veps_i}{\Edsgn}
    \label{eq:proof:unknown-ENR:Etot:Ni-choice}
     \\ &= \frac{x}{4}\left(\e^{\alpha L} - 1\right)\left(1 + \sqrt{1 + \frac{4\e^{\alpha\left(L + 1\right)}}{\left(1 - \e^{\alpha L}\right)^2}}\right)\frac{\e^{-2\alpha}}{1 - \e^{-\alpha}}
     \col{}{\\ &\qquad\qquad} + \half \left(\frac{\e^\alpha}{x}\right)^{L - 1}  + \sum_{i=2}^{\jmath} \frac{\veps_i}{\Edsgn} ,
    \label{eq:proof:unknown-ENR:Etot:define-x}
    \end{align}
    \end{subequations}
    where 
    \eqref{eq:proof:unknown-ENR:Etot:substitute} follows from \eqref{eq:FirstLayerEnergy} and \eqref{eq:Energy_Linear_Final},
    in \eqref{eq:proof:unknown-ENR:Etot:Ni-choice} we use the choice $N_i = \Delta \e^{- \alpha \left(i - 1\right)}$ for the noise levels for some positive parameters $\alpha$ and $\Delta$,
    and \eqref{eq:proof:unknown-ENR:Etot:define-x} holds by defining $x \triangleq \Delta / \Edsgn$.
    
    Finally, by optimizing over the parameters $\alpha$ and $x$, taking a large enough $M$, and taking $k$ to infinity, we arrive at the desired result.

    For the particular case of a quadratically decaying profile ($L=2$), numerically optimizing \eqref{eq:proof:unknown-ENR:Etot:define-x} over $\alpha$ and $x$ yields~\eqref{eq:MinimalEnergy_Linear_quad}.


\section{Proof of \thmref{thm:quadratic_profile}}
\label{app:proof:thm:quadratic_profile}

To prove \thmref{thm:quadratic_profile}, 
we will make use of the following non-uniform variant of the Berry--Esseen theorem, which is a weakened (yet more compact) form of a result due to Petrov.

\begin{thm}[\!\!{\cite{Petrov_BerryEssenPDF}, \cite[Ch.~VII, Thm.~17]{Petrov:SumsOfIndependentRVs:Book}}]
\label{thm:Petrov:CLT}
    Let $\{x_i | i \in \nats \}$ be an \iid\ sequence of RVs with zero mean and unit variance, and denote 
    $s_n \triangleq \frac{1}{\sqrt{n}} \sum_{i=1}^n x_i$. 
    Assume that $\E{|x_1|^\nu} < \infty$ for some $\nu > 2$, and that $x_1$ has a bounded p.d.f.\ 
    Then, the p.d.f.\ of $s_n$, denoted by $f_n$, satisfies 
    \begin{align} 
        \left| f_n(a) - f_G(a) \right| &< \frac{A_\nu}{\sqrt{n} \cdot \left( 1 + |a|^\nu \right)}, & \forall a &\in \reals ,
    \end{align} 
    for some $A_\nu < \infty$, where $f_G$ is the standard Gaussian p.d.f.
\end{thm}

Furthermore, as discussed in Rems.~\ref{rem:Interleaving}--\ref{rem:induced-channel},
in each layer (and specifically in the PPM layer) we effectively have an additive noisy channel whose noise distribution approaches a Gaussian distribution (due to the Gaussianization and the interleaving). For all the linear layers, \lemref{lem:error:semi-norm-ergodic} allow us to use the MLM results of \secref{s: MLM}. However, for the PPM layer, instead of a combination of Gaussian vectors and dithers (as treated in \lemref{lem:error:semi-norm-ergodic}) the actual noise contains also terms that are induced by the PPM scheme. We will use the following lemma, which is proved in \appref{app:SNR_Perturbation}, to claim that the effect of these terms on the performance of the MLM scheme can be made arbitrarily small, by taking the dimension $k$ to be large enough.

\begin{lem}
\label{lem:Pertub_SNE}
    Let $x^k$ be a sequence in $k$ of SNE vectors with second moment $r_k$ such that $\lim_{k \to \infty} r_k = r$, and let $\hx^k$ be a corresponding sequence in $k$ of vectors with identically distributed entries such that: 
    \begin{itemize}
        \item 
        The distance between the p.d.f.\ of $x_1$ and that of $\hx_1$ is bounded from above by 
            \begin{align}
                \label{eq:ProbBound_SNE_Prf}
                 \abs{f_{\hx_1}(t) - f_{x_1}(t)} &\leq \frac{C(t)}{\sqrt{k}} 
                & \forall t > 0 ,
            \end{align}
            where $C(t) = o\left (1/t^2 \right)$,\footnote{$f(t)=o(g(t))$ means that $\lim_{t \to \infty} \frac{f(t)}{g(t)} = 0$.} and where $x_1$ and $\hx_1$ denote the first entries of the vectors $x^k$ and $\hx^k$, respectively.
        \item 
            The correlation between any two squared  entries within $\hx^k$ decays to zero with $k$, \viz., 
            \begin{align}
            \label{eq:Slutsky}
                \lim_{k \to \infty} \cov{\hx^2_i}{\hx^2_j} = 0 ,
            \end{align}
            where $\cov{A}{B} \triangleq \E{AB} - \E{A}\E{B}$ denotes the covariance of $A$ and $B$.
    \end{itemize}
    
    Then, for all $\epsilon, \delta > 0$,
    there exists $k_0 \in \nats$ such that  
    \begin{align}
        P\left(\frac{1}{k}\norm{\hx^k}^{2} > r + \delta \right) \leq \epsilon 
    \end{align}
    for all $k > k_0$, 
    namely, the sequence $\left\{\hx^k \middle| k \in \nats \right\}$ is a sequence of SNE vectors.
\end{lem}

We will now prove \thmref{thm:quadratic_profile}. We note that the following analysis is based on the interleaving and Gaussianization blocks as they appear in the full description of the scheme in \appref{app:Full_M_Layer}.

\begin{IEEEproof}[Proof of \thmref{thm:quadratic_profile}]
    We will now derive the parameters that achieve a quadratic profile ($L=2$) and $\Edsgn$ in \eqref{eq:Profile_Polynom} 
    for all $N > \Nmin$ for a given $\Nmin > 0$.
    
    We choose $\mH_i = I_B$ for the linear layers---layers $i = \{1, 2, \ldots, M-1\}$.
    Consequently, the analysis for the first $M-1$ layers of the proof of \thmref{thm:linear_MLM} carries over to this scheme as well.

    Consider now the last layer---layer $M$.
    Following Feder and Ingber \cite{Feder-Ingber:Patent:Hadamard2012}, and 
    Hadad and Erez \cite{HadadErez:Dithered:Gaussianize:SP2016}, 
    we use a $B$-dimensional Walsh--Hadamard matrix $\mH_M$.

    Now, if $N \in \left( N_{M-1}, N_M \right]$, the receiver uses the last layer to improve the source estimates while viewing the estimates resulting from the previous layer, $\{\hx_{M - 1}^k(1), \ldots, \hx_{M - 1}^k(B)\}$, as SI with mean power $D_{M - 1}(N_{M})$ \eqref{eq:proof:unknown-ENR:Di}.

    By \lemref{lem:f_dither<f_G}, all the moments of 
 all the entries of $m_{M}^k(b)$ exist and are finite for all $b$.
    Thus, by \thmref{thm:Petrov:CLT}, and since $m_M^k(1), m_M^k(2),\ldots, m_M^k(B^k)$ are i.i.d.,
    the p.d.f.\ $f_\ell$ of $\tm_{M;(\ell, j)}(b)$ (it is the same for all $b$ and $j$ for a given $\ell$) satisfies 
    \begin{align}
        |f_\ell(a) - f_{G_\ell}(a)| < \frac{A_\nu}{\sqrt{B} \left( 1 + |a|^\nu \right)}
    \end{align}
    for all $\ell \in \{1, \ldots, k\}$, $b \in \{1, \ldots, B\}$, and $j \in \{1, 2, \ldots, B^{k-1} \}$, for all $\nu > 2$ 
    for some $A_\nu < \infty$, 
    where $f_{G_\ell}$ is the p.d.f.\ of a zero-mean Gaussian RV with the same variance as $\tm_{M;(\ell,j)}(b)$.
    
    By choosing some $\nu > 4$
    and applying \corref{cor:NearGaussianDist_UpperBound} to $\tm_{M;\ell}(b)$
    with $h(a) = A_\nu  / \left( 1 + |a|^\nu \right)$ and $\eps = 1/\sqrt{B}$, 
    the distortion bound of \thmref{thm:UpperBound_GaussianPrior} is attained up to a loss $C / \sqrt{B}$ for some constant $C < \infty$, where this loss can be made arbitrarily small by choosing a large enough $B$.
    
    We note that the interleaving makes the PPM transmitters operate over elements that are related to lattices of different sources. Thus, after deinterleaving, the correlation between different vector elements, as well as the correlation between their squares, vanishes as $k \to \infty$. Furthermore, the per-element variance is bounded from above by quantity that approaches (as $k \to \infty$) the PPM performance bound of \thmref{thm:UpperBound_GaussianPrior}. Thus, by \lemref{lem:Pertub_SNE}, the resulting effective noise vector $z_\eff^k = \hm_{M}^k (b) - m_{M}^k (b)$ is SNE (recall \defnref{def:SemiNorm}).

    We note that $z_\eff^k(b)$ is correlated with $m_{M}^k (b)$; nevertheless, by 
    \corref{cor:MLM:correlated_params} with parameters $\talpha = \alpha_c = \alpha_s = 1$ 
    the distortion of $\hx_{M}^k$ is bounded from above by 
    \begin{align}
        \label{eq:Distortion_PPMLayer}
        D_{M}(N) \leq \frac{D_{M - 1}(N_{M})}{\SDR_{M}\left(N\right)} + \eps_{M}, 
    \end{align}
    where $\eps_{M}$ subsumes the aforementioned losses that all go to zero with $k$, and $\SDR_{M}(N)$ is the SDR of the analog PPM scheme for a noise power $N$ of \thmref{thm:UpperBound_GaussianPrior}.

    The energy $E_{M}$ of the last layer is chosen to comply with the profile for $N < N_{M}$: 
    \begin{align}
    \label{eq:MinimumSDR_SquaredProfile}
        D_{M}\left(N\right) &\leq \mathcal{F}\left(N\right) & \forall N < N_{M}.
    \end{align}
    Combining \eqref{eq:Distortion_PPMLayer} and the contribution of the first $M - 1$ layers, given by  \eqref{eq:proof:unknown-ENR:Etot:Ni-choice} with summation from $1$ to $M - 1$. By numerically optimizing the resulting term over the number of layers $M$, the PPM pulse width $\beta$ and the energy layers $\left\{E_i\right\}_{i=1}^{M}$ we obtain that $M = 7$, $\beta = 0.9$ and the layer energies  
    $E_1 \approx 0.8480 \Edsgn, E_2 \approx 0.4893 \Edsgn, E_3 \approx 0.2823 \Edsgn, E_4 \approx 0.1629 \Edsgn$, $E_5 \approx 0.094 \Edsgn$, $E_6 \approx 0.0542 \Edsgn$, $E_7 \approx 0.0313 \Edsgn$
    yields \eqref{eq:Emin:analog_ppm_mlm}. 
\end{IEEEproof}


\section{Proof of \lemref{lem:error:semi-norm-ergodic}}
\label{app:error:semi-norm-ergodic}

Since $\Lambda^{(k)}$ is assumed to be a sequence that is good for channel coding, 
\begin{align}
\label{eq:lattice-error}
    \lim_{k \to \infty} \PR{x^k - \hx^k = e^k} = 1,
\end{align}
where $e^k \triangleq (1 - \alpha_s) q^k - \frac{\alpha_s \alpha_c}{\beta} z^k + \frac{\alpha_s \left(1 - \alpha_c\right)}{\beta} m^k$; equivalently, 
for any $\eps_1 > 0$, however small, there exists $k_1 \in \nats$, such that for all $k > k_1$, 
\begin{align}
    \PR{x^k - \hx^k \neq e^k} < \eps_1. 
\end{align}

Note now that $e^k$ equals a linear combination of independent Gaussian vectors---which amounts to a Gaussian vector---and dither vectors. 
Hence, by \cite[Th.~3]{OrdentlichErez_LatticeRobustness}, 
the sequence in $k$ of vectors $e^k$ is SNE, namely, 
for any $\delta, \eps_2 > 0$, however small, there exists $k_2 \in \nats$, such that for all $k > k_1$, 
\begin{align}
    \PR{ \frac{1}{k}\E{\norm{e^k}^2} >  (1 + \delta) \sigma_e^2 } \leq \eps_2 .
\end{align}

Now let $\eps > 0$, however small and choose $\eps_1 = \eps_2 = \eps / 2$, and $k_0 = \max\{k_1, k_2\}$. Then, by the union bound, for all $k > k_0$, 
\begin{align}
    \PR{ \frac{1}{k}\E{\norm{x^k - \hx^k}^2} >  (1 + \delta) \sigma_{x - \hx}^2 } \leq \eps .
\end{align}


\section{Proof of \lemref{lem:Pertub_SNE}}
\label{app:SNR_Perturbation}
    By \eqref{eq:ProbBound_SNE_Prf}, 
    \begin{align}
        \label{eq:SecondMoment_xHat}
        \abs{ \E{\hx^2_1} - \E{x_1^2} } 
        = \abs{ \int_{t=\infty}^{\infty} t^2 \left[  f_{\hx_1}(t) - f_{x_1}(t) \right] dt }
        \leq \frac{G}{\sqrt{k}}
    \end{align}
    where $G$ is a \textit{finite} constant, since $C(t) = o(t)$, that depends on $C(t)$, and $\E{x_1^2} = r_k$. 
    Using second-moment ergodicity of the entries of $x^k$, which holds in the limit of $k \to \infty$ \cite[Thm.~12.1]{PapoulisBook:4thEd} by \eqref{eq:Slutsky} and recalling that $\lim_{k \to \infty} r_k = r$
    concludes the proof. 

\bibliographystyle{IEEEtran}
\bibliography{myBib}

\begin{thebibliography}{10}
\providecommand{\url}[1]{#1}
\csname url@samestyle\endcsname
\providecommand{\newblock}{\relax}
\providecommand{\bibinfo}[2]{#2}
\providecommand{\BIBentrySTDinterwordspacing}{\spaceskip=0pt\relax}
\providecommand{\BIBentryALTinterwordstretchfactor}{4}
\providecommand{\BIBentryALTinterwordspacing}{\spaceskip=\fontdimen2\font plus
\BIBentryALTinterwordstretchfactor\fontdimen3\font minus
  \fontdimen4\font\relax}
\providecommand{\BIBforeignlanguage}[2]{{%
\expandafter\ifx\csname l@#1\endcsname\relax
\typeout{** WARNING: IEEEtran.bst: No hyphenation pattern has been}%
\typeout{** loaded for the language `#1'. Using the pattern for}%
\typeout{** the default language instead.}%
\else
\language=\csname l@#1\endcsname
\fi
#2}}
\providecommand{\BIBdecl}{\relax}
\BIBdecl

\bibitem{CoverBook2Edition}
T.~M. Cover and J.~A. Thomas, \emph{Elements of Information Theory, Second
  Edition}.\hskip 1em plus 0.5em minus 0.4em\relax New York: Wiley, 2006.

\bibitem{ElGamalKimBook}
A.~{El Gamal} and Y.-H. Kim, \emph{Network Information Theory}.\hskip 1em plus
  0.5em minus 0.4em\relax Cambridge University Press, 2011.

\bibitem{Shannon59:RDF}
C.~E. Shannon, ``Coding theorems for a discrete source with a fidelity
  criterion,'' in \emph{Institute of Radio Engineers, International Convention
  Record}, vol.~7, 1959, pp. 142--163.

\bibitem{KokenTuncel}
E.~{Köken} and E.~{Tuncel}, ``On minimum energy for robust {Gaussian} joint
  source--channel coding with a distortion--noise profile,'' in
  \emph{Proceedings of the IEEE International Symposium on Information Theory
  (ISIT)}, Aachen, Germany, 2017, pp. 1668--1672.

\bibitem{EquitzCover91}
W.~H.~R. Equitz and T.~M. Cover, ``Successive refinement of information,''
  \emph{IEEE Transactions on Information Theory}, vol.~37, no.~2, pp. 851--857,
  Mar. 1991.

\bibitem{Santhi-Vardy:JSCC:optimal-slope:ISIT}
N.~Santhi and A.~Vardy, ``Analog codes on graphs,'' in \emph{Proceedings of the
  IEEE International Symposium on Information Theory (ISIT)}, Yokohama, Japan,
  2003, p.~13.

\bibitem{Santhi-Vardy:JSCC:optimal-slope:Arxiv}
------, ``Analog codes on graphs,'' \emph{arXiv preprint cs/0608086}, 2006.

\bibitem{Bhattad-Narayanan:JSCC:optimal-slope}
K.~Bhattad and K.~R. Narayanan, ``A note on the rate of decay of mean-squared
  error with snr for the awgn channel,'' \emph{IEEE Transactions on Information
  Theory}, vol.~56, no.~1, pp. 332--335, 2010.

\bibitem{MittalPhamdo}
U.~Mittal and N.~Phamdo, ``Hybrid digital-analog ({HDA}) joint source--channel
  codes for broadcasting and robust communications,'' \emph{IEEE Transactions
  on Information Theory}, vol.~48, no.~5, pp. 1082--1102, May 2002.

\bibitem{Reznic}
Z.~Reznic, M.~Feder, and R.~Zamir, ``Distortion bounds for broadcasting with
  bandwidth expansion,'' \emph{IEEE Transactions on Information Theory},
  vol.~52, no.~8, pp. 3778--3788, Aug.~2006.

\bibitem{JointWZ-WDP}
Y.~Kochman and R.~Zamir, ``Joint {W}yner--{Z}iv/dirty-paper coding by
  modulo-lattice modulation,'' \emph{IEEE Transactions on Information Theory},
  vol.~55, pp. 4878--4899, Nov. 2009.

\bibitem{AnalogMatching}
------, ``Analog matching of colored sources to colored channels,'' \emph{IEEE
  Transactions on Information Theory}, vol.~57, no.~6, pp. 3180--3195, June
  2011.

\bibitem{ZamirBook}
R.~Zamir, \emph{Lattice Coding for Signals and Networks}.\hskip 1em plus 0.5em
  minus 0.4em\relax Cambridge: Cambridge University Press, 2014.

\bibitem{WynerZiv76}
A.~D. Wyner and J.~Ziv, ``The rate--distortion function for source coding with
  side information at the decoder,'' \emph{IEEE Transactions on Information
  Theory}, vol.~22, no.~1, pp. 1--10, Jan. 1976.

\bibitem{Wyner78}
A.~D. Wyner, ``The rate--distortion function for source coding with side
  information at the decoder---{II}: General sources,'' \emph{Information and
  Control}, vol.~38, pp. 60--80, 1978.

\bibitem{MinimumEnergyBound_Tuncel}
E.~{Köken} and E.~{Tuncel}, ``On minimum energy for robust {Gaussian} joint
  source--channel coding with a distortion--noise profile,'' in
  \emph{Proceedings of the IEEE International Symposium on Information Theory
  (ISIT)}, 2017, pp. 1668--1672.

\bibitem{baniasadi2020minimum}
M.~Baniasadi and E.~Tuncel, ``Minimum energy analysis for robust {Gaussian}
  joint source--channel coding with a square-law profile,'' in
  \emph{Proceedings of the IEEE International Symposium on Information Theory
  and Its Applications (ISITA)}, 2020, pp. 51--55.

\bibitem{Baniasadi-Koken-Tuncel:energy-limited-JSCC:universal:IT2022}
M.~Baniasadi, E.~K{\"o}ken, and E.~Tuncel, ``Minimum energy analysis for robust
  {Gaussian} joint source--channel coding with a distortion--noise profile,''
  \emph{IEEE Transactions on Information Theory}, vol.~68, no.~12, pp.
  7702--7713, Dec. 2022.

\bibitem{baniasadi202staircase}
\BIBentryALTinterwordspacing
M.~Baniasadi, ``Robust {Gaussian} joint source--channel coding with a staircase
  distortion--noise profile,'' \emph{CoRR}, 2020. [Online]. Available:
  \url{http://arxiv.org/abs/2001.09370}
\BIBentrySTDinterwordspacing

\bibitem{EnergyLimitedJSCC:Lev_Khina:Full}
O.~Lev and A.~Khina, ``Energy-limited joint source--channel coding via analog
  pulse position modulation,'' \emph{IEEE Transactions on Communications},
  vol.~70, no.~8, pp. 5140--5150, August 2022.

\bibitem{OrdentlichErez_LatticeRobustness}
O.~Ordentlich and U.~Erez, ``A simple proof for the existence of “good”
  pairs of nested lattices,'' \emph{IEEE Transactions on Information Theory},
  vol.~62, no.~8, pp. 4439--4453, 2016.

\bibitem{RematchAndForward_Full}
Y.~{Kochman}, A.~{Khina}, U.~{Erez}, and R.~{Zamir}, ``Rematch-and-forward:
  Joint source–channel coding for parallel relaying with spectral mismatch,''
  \emph{IEEE Transactions on Information Theory}, vol.~60, no.~1, pp. 605--622,
  2014.

\bibitem{ErezZamirAWGN}
U.~Erez and R.~Zamir, ``Achieving $\frac{1}{2}\log(1+{\SNR})$ on the {AWGN}
  channel with lattice encoding and decoding,'' \emph{IEEE Transactions on
  Information Theory}, vol.~50, no.~10, pp. 2293--2314, Oct. 2004.

\bibitem{WozencraftJacobsBook}
J.~M. Wozencraft and I.~M. Jacobs, \emph{Principles of Communication
  Engineering}.\hskip 1em plus 0.5em minus 0.4em\relax New York: John Wiley \&
  Sons, 1965.

\bibitem{ViterbiOmuraBook}
A.~J. Viterbi and J.~K. Omura, \emph{Principles of Digital Communication and
  Coding}.\hskip 1em plus 0.5em minus 0.4em\relax \!\!\!New York: McGraw-Hill,
  1979.

\bibitem{Wu-Verdu:MMSE:IT2011}
Y.~Wu and S.~Verd{\'u}, ``Functional properties of minimum mean-square error
  and mutual information,'' \emph{IEEE Transactions on Information Theory},
  vol.~58, no.~3, pp. 1289--1301, 2011.

\bibitem{Feder-Ingber:Patent:Hadamard2012}
M.~Feder and A.~Ingber, ``Method, device and system of reduced
  peak-to-average-ratio communication,'' U.S. Patent 11/971,934, Feb. 14, 2014.

\bibitem{HadadErez:Dithered:Gaussianize:SP2016}
R.~Hadad and U.~Erez, ``Dithered quantization via orthogonal transformations,''
  \emph{IEEE Transactions on Signal Processing}, vol.~64, no.~22, pp.
  5887--5900, 2016.

\bibitem{Asnani-Shomorony-Avestimehr-Weissman:WorstCaseCompression:Gaussianize:IT2015}
H.~Asnani, I.~Shomorony, A.~S. Avestimehr, and T.~Weissman, ``Network
  compression: Worst case analysis,'' \emph{IEEE Transactions on Information
  Theory}, vol.~61, no.~7, pp. 3980--3995, 2015.

\bibitem{NoWeissman:RDF:Extremes:IT2016}
A.~No and T.~Weissman, ``Rateless lossy compression via the extremes,''
  \emph{IEEE Transactions on Information Theory}, vol.~62, no.~10, pp.
  5484--5495, 2016.

\bibitem{baniasadi-tuncel-SI:ISIT2021}
M.~Baniasadi and E.~Tuncel, ``Robust {Gaussian JSCC} under the near-infinity
  bandwidth regime with side information at the receiver,'' in
  \emph{Proceedings of the IEEE International Symposium on Information Theory
  (ISIT)}, 2021.

\bibitem{baniasadi-tuncel:ZeroBW:ISIT2020}
------, ``Robust {Gaussian} joint source--channel coding under the near-zero
  bandwidth regime,'' in \emph{Proceedings of the IEEE International Symposium
  on Information Theory (ISIT)}, 2020, pp. 2474--2479.

\bibitem{Petrov_BerryEssenPDF}
V.~V. Petrov, ``On local limit theorems for sums of independent random
  variables,'' \emph{Theory of Probability \& Its Applications}, vol.~9, no.~2,
  pp. 312--320, 1964.

\bibitem{Petrov:SumsOfIndependentRVs:Book}
------, \emph{Sums of Independent Random Variables}.\hskip 1em plus 0.5em minus
  0.4em\relax New York: Springer-Verlag, 1975.

\bibitem{PapoulisBook:4thEd}
A.~Papoulis and S.~U. Pillai, \emph{Probability, random variables, and
  stochastic processes}, 4th~ed.\hskip 1em plus 0.5em minus 0.4em\relax Tata
  McGraw-Hill Education, 2002.

\end{thebibliography}
\end{document}